\documentclass[fp,twocolumn]{jpsj3}
\usepackage{txfonts}
\usepackage[english]{babel}
\usepackage{graphicx}
\graphicspath{{figures/}}
\usepackage{braket}
\usepackage{dcolumn}
\usepackage{amsfonts}
\usepackage{url}
\usepackage[table]{xcolor}
\usepackage{multirow}
\usepackage{placeins}
\usepackage[colorlinks=true,linkcolor=blue,citecolor=blue,urlcolor=blue]{hyperref}

\makeatletter
\providecommand{\ext@figure}{lof}
\providecommand{\ext@table}{lot}
\makeatother

\begin{document}

\title{Resonant Suppression of the \texorpdfstring{$\pi$}{pi}-Type Josephson Component in a Kitaev-Ladder Junction}
\author{Cheng-Rong Xie$^{1,2}$\thanks{Corresponding author: \url{xie.chengrong.s4@alumni.tohoku.ac.jp}}, Hiroki Tsuchiura$^{1}$, and Yukio Tanaka$^{3}$}
\inst{$^{1}$Department of Applied Physics, Tohoku University, Sendai 980-8579, Japan\\
$^{2}$Institute for Advanced Study, Shenzhen University, Shenzhen 518060, China\\
$^{3}$Department of Applied Physics, Nagoya University, Nagoya 464-8603, Japan}
\abst{We study how discrete normal-segment levels control equilibrium Josephson transport between two Kitaev-ladder superconductors. Away from resonance, detuning opens a selective-clamping window: inter-leg hybridization removes the phase dependence of a $0$-favoring low-energy ground-state contribution while retaining a competing $\pi$-favoring contribution, allowing a smooth negative first harmonic even for bulk-topologically trivial leads. At resonance and $\phi=\pi$, an exact collective-clamping identity makes both low-energy ground-state contributions phase independent above a threshold, although individual quasiparticle levels can remain dispersive. Finite-energy quasiparticles then produce a positive residual first harmonic in a controlled sweet-point regime. For fixed $0<\phi<\pi$, helicity mixing instead yields a positive resonant low-energy harmonic in the weak-coupling limit. Exact identities, fourth-order sign rules, and recursive Green-function calculations establish how resonant alignment changes the spectral contributions governing Josephson phase selection.}
\maketitle

\section{Introduction}
Topological superconducting junctions provide an ideal setting in which the Josephson current--phase relation (CPR) probes boundary quasiparticle states beyond the bulk spectrum alone.\cite{Golubov2004,Read2000,A.Yu.Kitaev2001,Fu2008,Tanaka2009,Lutchyn2010,Oreg2010,Tanaka2012,Alicea_2012,Beenakker2013,Elliott2015,Sato2016} Phase-sensitive Josephson transport and sign-changing couplings reflect unconventional pairing and boundary-state hybridization,\cite{Tanaka1996,Tanaka1997,SatoshiKashiwaya_2000,Kirtley2005} with Majorana systems attracting intense interest due to parity-protected crossings and fractional Josephson responses.\cite{Tanaka2024,Kwon2004,Lee2014,Fu2009,Jiang2011,Zhang2014,Badiane2011,vanHeck2011,Laroche2019} The Kitaev ladder extends the single-chain geometry by incorporating an inter-leg hopping $t_\perp$ and an inter-leg phase difference $\phi$, which introduce multiple end-mode degrees of freedom and competing tunneling channels.\cite{PhysRevB.89.174514,Maiellaro2018,Yan2020} Previous theory identified competition between end-state $\pi$ coupling and conventional interchain Josephson coupling.\cite{Nakosai2018} Kitaev-ladder junctions can display fractional, conventional $0$-, and $\pi$-type Josephson responses depending on the ladder parameters.\cite{Kanehira2023JosephsonEB} Related $\pi$-shifted junctions have also been studied in other topological-superconductor settings.\cite{PhysRevB.87.100506,PhysRevB.100.064505}

In mesoscopic superconducting weak links, resonant transmission through discrete levels can enhance Josephson transport.\cite{GlazmanMatveev1989,BeenakkervanHouten1992} Interaction- and impurity-driven $0$--$\pi$ reversals have also been investigated in quantum-dot and Yu--Shiba--Rusinov junctions.\cite{Rozhkov1999,Karan2022} Continuum and higher-energy quasiparticle states may additionally contribute to the equilibrium Josephson current.\cite{Krichevsky2000} For a single Kitaev chain in an S/N/S geometry, Sumita and Furusaki demonstrated that discrete levels in the normal segment mediate virtual cotunneling off resonance and substantially amplify the fractional Josephson current on resonance.\cite{PhysRevB.104.205431} In a Kitaev ladder, however, the transverse leg degree of freedom generates competing low-energy boundary branches. How discrete-level resonance in a finite normal segment affects this branch competition and governs the equilibrium phase selection remains an open question.

In this paper, we investigate an S/N/S junction connecting two semi-infinite Kitaev ladders via a finite normal segment. Combining recursive Green's function calculations, as in Refs. \citen{Sancho_1985,PhysRevB.55.5266,PhysRevB.87.024515}, with controlled analytical expansions, we find that discrete-level resonance can reverse the dominant Josephson harmonic by changing the spectral contributions that control equilibrium phase selection. Away from resonance, detuning opens a selective-clamping window whose exact width in the isolated-channel limit is $|\varepsilon|$: within this window, inter-leg hybridization suppresses the phase dependence of the $0$-favoring ground-state contribution while leaving the competing $\pi$-favoring contribution dispersive. This imbalance generates a negative first Josephson harmonic ($J_1<0$), yielding a smooth $\pi$-type response in the crossing-free regime even when the superconducting leads are bulk-topologically trivial.

Near a normal-level resonance, this pre-existing $\pi$ component is strongly suppressed. At $\phi=\pi$, collective clamping simultaneously renders the ground-state contributions of both low-energy branches independent of the Josephson phase. We establish an exact identity proving this cancellation for an arbitrary normal-segment length $L_M$. Finite-energy quasiparticle states then determine the residual coupling; at the Kitaev sweet point, its leading fourth-order first harmonic is positive ($J_1>0$) within the derived spectral window. Away from $\phi=\pi$, helicity mixing restores a positive low-energy harmonic under a controlled weak-coupling expansion. Recursive Green's function calculations exhibit the same resonance-induced suppression beyond the analytically controlled limits, including representative parameters away from the sweet point.

The paper is organized as follows. Section~\ref{Model and Formalism} introduces the S/N/S Kitaev-ladder model, the recursive Green's function method, and the resonance condition. Section~\ref{results} presents the CPR phase diagrams and resonance diagnostics. Section~\ref{current on} analyzes the branch-clamping mechanism, the finite-segment sign rules, and the general-$\phi$ expansion. Technical derivations and the isolated-ladder topology are provided in the Appendices.

\section{Model and Formalism}
\label{Model and Formalism}
\begin{figure}[htbp]
\centering
\includegraphics[width=0.48\textwidth]{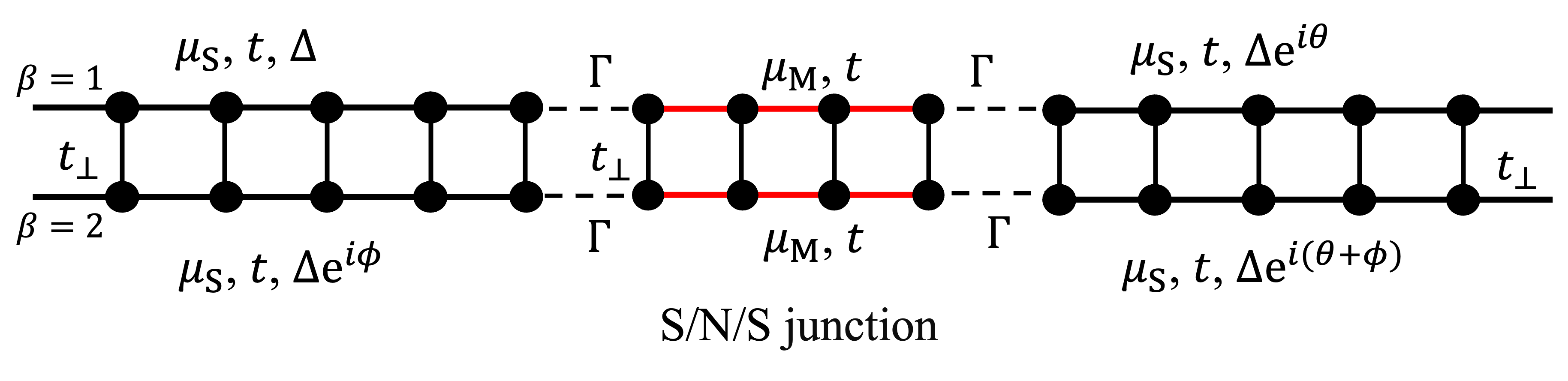}
\caption{\label{fig:kitaev ladder}Kitaev-ladder S/N/S junction.
Black and red segments denote the superconducting leads and the finite
normal region, respectively. The legs are labeled by $\beta=1,2$.
$t_\perp$ is the inter-leg hopping, $\Gamma$ the interface hopping,
$\phi$ the phase difference between the legs, and $\theta$ the phase
difference across the junction.}
\end{figure}

We consider a system consisting of two semi-infinite Kitaev ladders connected by a finite segment of normal conducting sites, as shown in Fig.~\ref{fig:kitaev ladder}. The Josephson current across the junction is calculated using the recursive Green's function method, which evaluates local Green's functions between sites.\cite{Sancho_1985, Yada2014, PhysRevB.55.5266, PhysRevB.87.024515, PhysRevB.101.024509}

The total Hamiltonian of the junction system is written as
\begin{equation}
\mathcal{H}
=
\mathcal{H}_{\mathrm{S_L}}
+
\mathcal{H}_{\mathrm{S_R}}
+
\mathcal{H}_{\mathrm{M}}
+
\mathcal{H}_{\Gamma},
\end{equation}
where $\mathcal{H}_{\mathrm{S_L}}$ ($\mathcal{H}_{\mathrm{S_R}}$) describes the left (right) semi-infinite superconducting Kitaev ladder,
$\mathcal{H}_{\mathrm{M}}$ the finite normal segment,
and $\mathcal{H}_{\Gamma}$ the coupling between the superconducting and normal regions.

Below we explicitly present the Hamiltonian for a single Kitaev chain and the ladder structure.
The normal segment is described by the same tight-binding Hamiltonian as the superconducting ladder, but with the pairing potential set to zero, i.e., $\Delta = 0$.
We denote the chemical potentials of the superconducting leads and the
normal segment by $\mu_{\rm S}$ and $\mu_{\rm M}$, respectively (the hopping
amplitudes are identical in all regions). Each superconducting lead is a Kitaev ladder of length $L_S$, with $L_S$ sites on each leg and $L_S\to\infty$ in the semi-infinite limit. We use $\mathrm{M}$ to label the middle normal segment. The phase diagrams of
Sec.~\ref{results} (Figs.~\ref{current-phase diagram},~\ref{fig:137})
use $\mu_{\rm S}=\mu_{\rm M}=\mu$. The resonance diagnostics of
Fig.~\ref{fig:resonance-diag}(a)--(d) fix $\mu_{\rm S}=0$
and scan $\mu_{\rm M}$ independently, isolating the normal-level detuning
with the lead structure unchanged.
The coupling between the superconducting (S) and normal (N) regions is introduced through the interfacial hopping Hamiltonian
\begin{equation}
\mathcal{H}_{\Gamma}
=
-\Gamma \sum_{\beta=1,2}
\Big(
c_{\beta\,\mathrm{L},L_S}^{\dagger} c_{\beta\,\mathrm{M},1}
+
c_{\beta\,\mathrm{M},L_\text{M}}^{\dagger} c_{\beta\,\mathrm{R},1}
+
\mathrm{h.c.}
\Big),
\end{equation}
where $\Gamma$ denotes the hopping amplitude at the S--N interfaces, and $\beta$ denotes the chain index.
The superconducting and normal regions are coupled only at the two interfaces,
while the bulk Hamiltonian remains unchanged. Here the indices are ordered as $(\text{chain index},\;\mathrm{region},\;\mathrm{site})$, 
where $\mathrm{L}$, $\mathrm{M}$, and $\mathrm{R}$ denote the left superconducting,
normal-metal, and right superconducting regions, respectively, and $L_\text{M}$ is the
length of the normal segment.
When comparing different normal-segment lengths $L_{\rm M}$, we keep the remaining model parameters fixed at the values specified for each comparison.

As the elementary building block, we use the standard Kitaev-chain Hamiltonian:\cite{A.Yu.Kitaev2001}
\begin{equation}
\begin{split}
\mathcal{H}_{\mathrm{chain}}
=& -\mu_S \sum_x^{L_S} \left(c_{x}^{\dagger} c_{x} - \frac{1}{2} \right)
    - t \sum_{x=1}^{L_S-1} \left(c_{x}^{\dagger} c_{x+1} + \text{h.c.} \right)\\
&\quad + \sum_{x=1}^{L_S-1} \left( \Delta c_{x}^{\dagger} c_{x+1}^{\dagger} + \text{h.c.} \right),
\end{split}
\label{eq:chain-H}
\end{equation}
where $\mu_S$ is the chemical potential of the superconducting chains, $t$ is the hopping amplitude, and $\Delta = |\Delta|$ is the $p$-wave superconducting pairing amplitude.

The two semi-infinite superconducting leads are Kitaev ladders
composed of two such chains coupled by an inter-leg hopping. The
corresponding lead Hamiltonian is
\begin{equation}\begin{aligned}
\mathcal{H}_{\mathrm{lad}}
&=-\mu_S \sum_{\beta,x}^{L_S}
\left(c_{\beta,x}^{\dagger} c_{\beta,x} - \tfrac{1}{2} \right)
-t_\perp \sum_x^{L_S}
\left( c_{1,x}^\dagger c_{2,x} + \text{h.c.} \right)
\\
&\quad
-t \sum_{\beta,x}^{L_S-1}
\left(c_{\beta,x}^{\dagger} c_{\beta,x+1} + \text{h.c.} \right)
\\
&\quad
+\sum_{\beta,x}^{L_S-1}
\left( \Delta_{\beta\alpha} c_{\beta,x}^{\dagger} c_{\beta,x+1}^{\dagger}
+ \text{h.c.} \right),
\label{eq:kitaev ladder}
\end{aligned}\end{equation}
where $\Delta_{\beta\alpha} = |\Delta| e^{-i(\theta_\alpha + \phi_\beta)}$
with $\phi_1=0$, $\phi_2=\phi$ (the interchain phase difference
$\phi\equiv\phi_2-\phi_1$) and $\theta_\alpha$ the phase of lead
$\alpha$, $\theta_R-\theta_L=\theta$ the phase difference across the
junction. The corresponding Nambu block matrices and the recursive
Green-function construction are collected in
Appendix~\ref{app:rgf}.

In addition to the general model definition, it is useful to identify a
special parameter regime associated with an on-resonance condition in the
normal segment, which plays an important role in the Josephson coupling
discussed below.

In the Kitaev ladder model considered here, the finite normal segment
exhibits discrete energy levels given by
\begin{equation}
\begin{aligned}
\epsilon_{q,\eta}^{\mathrm{M}}
&= \mu_{\rm M} + \eta\, t_\perp
+ 2t \cos\left( \frac{\pi q}{L_{\rm M} + 1} \right),\\
&\qquad (q = 1, \dots, L_{\rm M},\; \eta=\pm),
\end{aligned}
\label{noraml energy}
\end{equation}
where the superscript $\mathrm{M}$ denotes the normal segment and the
$\eta\, t_\perp$ term originates from the bonding/antibonding
splitting due to the interchain coupling.  The normal-sector spectrum (corresponding to $\Delta=0$ with open boundary conditions) can be obtained via generalized diagonalization of banded Toeplitz matrices.\cite{PhysRevLett.117.076804,Cobanera_2017,PhysRevB.96.195133,PhysRevB.98.245423}
We define the on-resonance condition as the situation in which one of these
levels of the normal segment aligns with the hybridized end mode of the
same branch:\cite{PhysRevB.104.205431}
\begin{equation}
\epsilon_{q,\eta}^{\mathrm{M}} = \eta\, t_\perp.
\label{on resonance condition}
\end{equation}
Equation~\eqref{on resonance condition} defines the bare resonance
reference, corresponding to the alignment at $\phi=\pi$ in the
decoupled-interface limit. At a general inter-leg phase, the end-mode
energy is already shifted at $\Gamma=0$. For weak inter-leg hopping,
$E_{\mathrm{end},\eta}\simeq\eta t_\perp\sin(\phi/2)$, leading to a
displacement of the spectral alignment from the bare reference. Finite
interface coupling can introduce further corrections. We distinguish
these effects in the numerical analysis of Sec.~\ref{results}.

\paragraph{Josephson harmonics and CPR classification.}
Here $E_{\mathrm{GS}}(\theta)$ denotes the instantaneous equilibrium
BdG ground-state energy, obtained by occupying the negative-energy
states at each $\theta$,
\begin{equation}
E_{\mathrm{GS}}(\theta) = -\frac{1}{2}\sum_{E_a\geq0} E_a(\theta),
\label{eq:gs-convention}
\end{equation}
and the Josephson harmonics are its Fourier coefficients,
\begin{equation}
J_n=-\frac{1}{\pi}\int_0^{2\pi} E_{\mathrm{GS}}(\theta)\cos(n\theta)\,d\theta.
\label{eq:jn-def}
\end{equation}
Thus $E_{\mathrm{GS}}=-\sum_nJ_n\cos(n\theta)$ up to a constant and
$I(\theta)=(2e/\hbar)\partial_\theta E_{\mathrm{GS}}$. We write the
energy-normalized current-phase relation as
\begin{equation}
J(\theta)\equiv\frac{\hbar}{2e}I(\theta)
=
\partial_\theta E_{\mathrm{GS}}(\theta)
=
\sum_{n\ge1}nJ_n\sin(n\theta),
\label{eq:Jtheta-def}
\end{equation}
so that figure axes labeled $J(\theta)/t$ are dimensionless.

A zero-energy parity crossing appears as a cusp in $E_{\mathrm{GS}}$ and
a discontinuity in the equilibrium CPR. While adiabatically tracking an isolated,
parity-conserved Majorana Andreev level yields the characteristic
$4\pi$-periodic dispersion,\cite{Fu2009,Jiang2011,vanHeck2011} the
equilibrium ground-state energy $E_{\mathrm{GS}}(\theta)$ in Eq.~\eqref{eq:gs-convention}
minimizes over occupied quasiparticle states and is strictly $2\pi$-periodic.
We use the latter throughout, as appropriate for the
equilibrium Josephson response studied here.

For smooth, crossing-free CPRs the sign of $J_1$ distinguishes
$0$-type and $\pi$-type responses. In the presence of
parity-switching crossings, $J_1$ is used only as a measure of the
corresponding harmonic component and does not by itself classify the
junction as a conventional $0$ or $\pi$ junction. We then state the
CPR character (e.g.\ mixed topological--$\pi$) separately.

\section{Numerical Results}
\label{results}
\begin{figure*}[!t]
\centering
\includegraphics[width=\textwidth]
{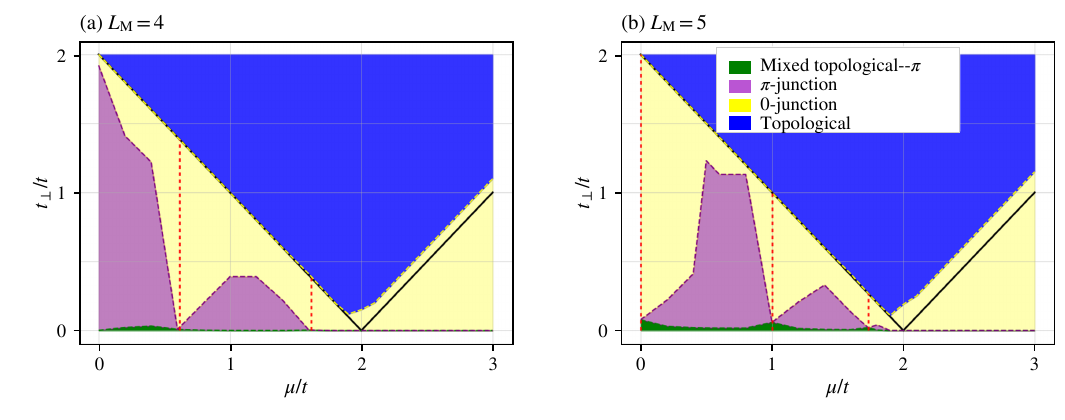}
\caption{CPR phase diagrams for normal segments of different lengths: (a) $L_{\rm M} = 4$ and (b) $L_{\rm M}=5$, with $\Delta=0.6t$. Colored regions represent distinct junction behaviors: topological junction (blue), $0$-junction (yellow), $\pi$-junction (purple), and mixed topological--$\pi$ regime (green). Black solid lines indicate the topological phase boundaries. Red dashed lines indicate the \emph{bare} resonance lines of Eq.~\eqref{on resonance condition} (the end-mode estimate gives a dressed-resonance shift of magnitude $t_\perp[1-\sin(\phi/2)]$, Sec.~\ref{results}). For $L_{\rm M}=4$, the bare lines occur at $|\mu/t|=0.618$ and $1.618$, as seen in (a). 
For $L_{\rm M}=5$, the resonance points shift to $|\mu/t|=0$, $1$, and $\sqrt3\simeq1.732$, as shown in (b). } \label{current-phase diagram}
\end{figure*}

In this section, we investigate the current-phase relations of the Kitaev ladder system with interchain phase difference $\phi=\pi/3$. Numerical parameters (in units of $t$) are $\Delta=0.6t$, interfacial hopping $\Gamma=0.1t$, temperature $T=0.0001t$ (ground-state limit, evaluated on the Matsubara axis with the cutoff increased until convergence), and normal-segment lengths $L_{\rm M}=4$ or $5$. Currents are normalized by $2et/\hbar$. We use the CPR classification to trace how the $\pi$-type component develops off resonance and is suppressed near the normal-level resonances of Sec.~\ref{Model and Formalism}.

Figure~\ref{current-phase diagram} summarizes the junction behavior in
the $(\mu,t_\perp)$ plane. Black solid lines mark the bulk topological
boundaries of the leads, whereas red dashed lines mark the bare
normal-level resonances. Their distinct trajectories separate changes
of the lead topology from changes caused by the finite normal segment.
The isolated-ladder classification is collected in
Appendix~\ref{Majorana zero modes on the Kitaev ladder}.

We call a CPR topological when it contains discontinuities associated
with parity-switching level crossings. Smooth responses are classified
by the first harmonic of Eq.~\eqref{eq:jn-def}: $J_1>0$ is $0$-type and
$J_1<0$ is $\pi$-type. In the mixed region a negative first harmonic
coexists with branch discontinuities.

Figure~\ref{current-phase diagram} reveals a direct relation between the $\pi$-type regions and the normal-level resonance lines. Away from the resonance lines, a sizable $\pi$-junction region appears even in parameter regimes where the bulk Kitaev ladder is topologically trivial. The $\pi$-type response in this bulk-trivial regime reflects the junction-level competition between end-mode branches reorganized by inter-leg coupling.

By contrast, near the red dashed resonance lines, the $\pi$-junction regions are strongly reduced or absent. Resonant tunneling through the discrete normal level reorganizes the low-energy branches that contribute to the Josephson current, thereby suppressing the off-resonant $\pi$-type component. At $\phi=\pi$, above the resonant threshold, the exact identity derived in Sec.~\ref{low-energy-picture} establishes that the ground-state contributions of both resonant low-energy branches become independent of the Josephson phase, as developed in Sec.~\ref{sec:clamping}. The shift of the resonance lines with $L_{\rm M}$ confirms that this suppression originates from the discrete level structure of the finite normal segment.

We first examine the off-resonant CPR evolution and then turn to the resonance diagnostics.

\subsection{Off-resonance case}

\begin{figure}[htbp]
\centering
\includegraphics[width=\columnwidth]{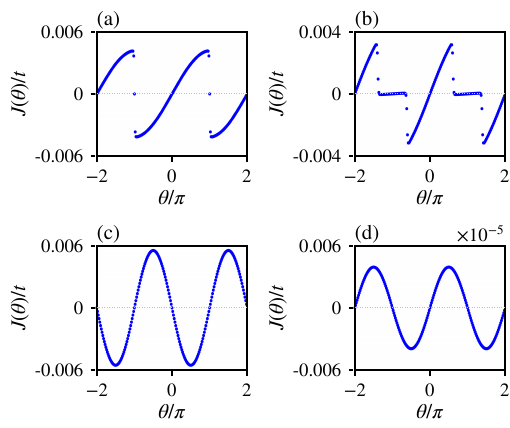}
\caption{
Current-phase relations for a normal segment of length
$L_{\rm M}=4$ and chemical potentials $\mu_{\rm S}/t=\mu_{\rm M}/t=0$,
with $\Delta/t=0.6$ and $\phi=\pi/3$.
Panels (a)-(d) correspond to
$t_\perp/t=0$, $0.01$, $0.1$, and $1.99$, respectively.
}
\label{fig:137}
\end{figure}

To identify the off-resonant mechanism behind the $\pi$-junction component, we examine representative CPRs along selected vertical cuts in the $(\mu,t_\perp)$ plane. We focus on $\mu/t=0$, which is away from the resonance lines for $L_{\rm M}=4$. The analytically tractable $\phi=\pi$ limit exposes the suppression of one low-energy ground-state contribution that produces the same negative-harmonic structure observed numerically away from resonance (Sec.~\ref{sec:clamping}).

At $t_\perp=0$ [Fig.~\ref{fig:137}(a)] the legs decouple into two
Kitaev chains. The sharp jumps at $\theta=\pi$ arise from
parity-switching zero-energy crossings and give the familiar
topological CPR.\cite{Kwon2004,Tanaka2024}

In Fig.~\ref{fig:137}(b), when $t_\perp \ll t$, the ladder enters a mixed-junction regime. 
The weak interchain coupling lifts the MZM degeneracy at $\theta=\pi$, 
leading to a hybridized level structure in which a discontinuity remains near $\theta=\pi$, 
while the local slope of the CPR changes sign compared to the purely topological case. 
As $t_\perp$ increases, Fig.~\ref{fig:137}(c) shows that the CPR becomes smoother, 
with its peak shifted toward $\theta=3\pi/2$, indicating the dominance of $\pi$-junction behavior. 
The current amplitude decreases as the hybridization increases.

When $t_\perp$ approaches $2t$, as shown in Fig.~\ref{fig:137}(d), the CPR further evolves into a sinusoidal form, characteristic of conventional $0$-junctions. The end-mode-mediated $\pi$-type contribution weakens as $t_\perp$ approaches the intrinsic superconducting scale, and the junction crosses back to a conventional $0$-type response. This re-entrant $0$-type behavior at large $t_\perp$ is a distinct strong-coupling change of the spectrum, outside the resonance mechanism studied below.

\subsection{On-resonance case}
We now examine parameter sets near the bare resonance lines highlighted in Fig.~\ref{current-phase diagram}(b). Near resonance, spectral alignment reorganizes the low-energy tunneling subspace and suppresses the $\pi$-type component of the Josephson coupling. Representative resonant CPRs are collected in Appendix~\ref{app:rep-cpr}.

\begin{figure*}[t]
\centering
\includegraphics[width=\textwidth]{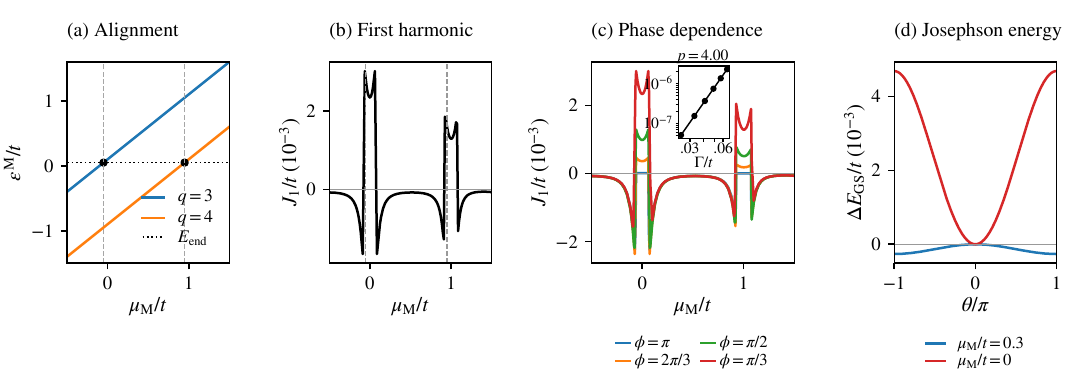}
\caption{Resonance diagnostics and Josephson energy ($L_{\rm M}=5$, $t_\perp=0.1t$, $\Gamma=0.1t$, $\Delta=0.6t$, $\mu_{\rm S}=0$).
(a) Normal-segment levels $\epsilon_{q,+}^{\mathrm{M}}$ for $q=3,4$
($\eta=+1$ branch of Eq.~\eqref{noraml energy}) vs the
end-mode energy $E_{\rm end}=0.049t$ at $\phi=\pi/3$. The black dots
mark the spectral-alignment estimates $\epsilon_{q,+}^{\mathrm{M}}=E_{\rm end}$ at
$\mu_{\rm M}\simeq-0.05$ ($q=3$) and $\mu_{\rm M}\simeq0.95$ ($q=4$).
(b) The first harmonic $J_1(\mu_{\rm M})$ becomes $0$-type in windows
appearing in the vicinity of these alignments and is $\pi$-type between and away from
them, so the suppression is tied to the spectral alignment.
The dashed lines mark the spectral-alignment estimates of panel (a).
They serve as guides to the resonance region rather than exact finite-$\Gamma$ resonance positions.
(c) The resonant $0$-type windows persist as the inter-leg phase is
varied from $\phi=\pi$ to $\phi=\pi/3$. Inset: $|J_1|$ versus $\Gamma$
for $\Gamma/t=0.025$--$0.065$, with $t_\perp=0.1t$ and
$\phi=\pi$, $\mu_{\rm M}=0$, so that $t_\perp>\sqrt2\,\Gamma$ throughout
the fitted range (distinct from the main-panel parameters
$t_\perp=\Gamma=0.1t$). The fitted exponent $p\simeq4$ is
consistent with the analytic absence of a phase-dependent
$O(\Gamma^2)$ term and with a residual equilibrium harmonic of
$O(\Gamma^4)$.
(d) Josephson energy $\Delta E_{\rm GS}=E_{\rm GS}(\theta)-E_{\rm GS}(0)$
at $\phi=\pi/3$: the off-resonant case ($\mu_{\rm M}/t=0.3$, blue)
has its minimum at $\theta=\pm\pi$ ($\pi$-type), whereas the near-resonant
reference ($\mu_{\rm M}=0$, red) has its minimum at $\theta=0$ ($0$-type).
The $q=3$ spectral alignment in panel (a) is at
$\mu_{\rm M}/t\simeq-0.05$.}
\label{fig:resonance-diag}
\end{figure*}

Equation~\eqref{on resonance condition} defines the bare resonance grid in the decoupled limit. Away from $\phi=\pi$, the relevant spectral alignment is shifted even in the decoupled-interface limit, because the hybridized ladder end-mode energy depends on the inter-leg phase. For finite $\Gamma$, additional shifts may occur. The alignment condition is
\begin{equation}
\epsilon^M_{q,\eta} = E_{{\rm end},\eta}(\phi),
\label{eq:spectral-alignment}
\end{equation}
which carries a branch index $\eta$. For the parameters used below,
$E_{{\rm end},\eta}(\phi)\simeq\eta t_\perp\sin(\phi/2)$, so the
alignments are shifted from the bare resonance lines by
$\eta t_\perp[\sin(\phi/2)-1]$, of magnitude
$t_\perp[1-\sin(\phi/2)]$. Figure~\ref{fig:resonance-diag}
resolves this shift directly.

Figure~\ref{fig:resonance-diag} thus links normal-level alignment to
a change in the equilibrium phase selected by the junction. The approximately
quartic interface-coupling dependence further points to a residual
Josephson contribution that is analyzed microscopically in
Sec.~\ref{current on}.

For the resonant one-rung junction ($L_{\rm M}=1$, $\mu_{\rm M}=0$),
the full numerical calculation at $\phi=\pi$, $\Delta=0.6t$, and
$t_\perp=0.1t$ gives $J_1\simeq+0.150\,\Gamma^4/t^3$ in the
weak-interface, zero-energy-crossing-free regime. The positive sign and quartic scaling agree with the sweet-point
prediction in Sec.~\ref{normallevels}; the numerical coefficient
$0.150$ differs from the sweet-point value $5/64$ because this
calculation uses $\Delta=0.6t$. The combined ground-state contributions of the resonant
low-energy branches are independent of the Josephson phase, whereas
finite-energy quasiparticle states carry the positive residual
Josephson coupling (Sec.~\ref{current on}).

\section{Spectral Origin of Resonance-Controlled Josephson Coupling}\label{current on}

The numerical results demonstrate that the first Josephson harmonic
changes sign near discrete normal-level resonances. To identify the
origin of this resonance-dependent coupling, we combine the exact
normal-mode propagator of the finite normal segment with the
superconducting boundary Green functions. This formulation retains
coherent contributions from all longitudinal modes and provides a
common starting point for the detuned and resonant regimes.
Additional helicity symmetry at $\phi=\pi$ allows the phase-dependent
spectrum to be separated exactly, while controlled weak-coupling
expansions determine the resulting Josephson harmonics. We first
identify the coherent propagation channels at general inter-leg
phase, then exploit the $\phi=\pi$ helicity structure, and finally
recover the spectral contributions responsible for the negative
off-resonant and positive resonant harmonics.

\subsection{General spectral formulation}\label{sec:general-spectral}
The inter-leg phase difference controls how the two legs participate in
Josephson transport. In the gauge where the intraleg pairing is real,
the rung coupling decomposes exactly as
\begin{equation}\begin{aligned}
&-t_\perp\left(e^{i\phi/2}c_1^\dagger c_2+e^{-i\phi/2}c_2^\dagger c_1\right)\\
&\quad=
-it_\perp\cos\frac{\delta}{2}\,
(c_1^\dagger c_2-c_2^\dagger c_1)
+t_\perp\sin\frac{\delta}{2}\,
(c_1^\dagger c_2+c_2^\dagger c_1),
\end{aligned}\label{eq:trung-delta}\end{equation}
with $\delta=\phi-\pi$, matching the Majorana form
Eq.~\eqref{eq:helicity-decomp} of Appendix~\ref{app:general-phi}.
Here helicity refers to the eigenvalue of $\sigma_y$ in leg space,
which becomes a conserved quantum number at $\phi=\pi$. The first
term is helicity preserving and the second mixes the two helicity
sectors. Only at $\phi=\pi$ ($\delta=0$) is helicity conserved
exactly. The local expansion about that point is deferred to
Sec.~\ref{robustness}.

The finite uniform normal chain has eigenfunctions
$\psi_q(j)=\sqrt{2/(L_M+1)}\sin(\pi qj/(L_M+1))$ with longitudinal
normal-mode energies
$\xi_q=\mu_{\rm M}+2t\cos k_q$, $k_q=\pi q/(L_M+1)$. These are related
to the ladder levels of Eq.~\eqref{noraml energy} by
$\epsilon_{q,\eta}^{\mathrm{M}}=\xi_q+\eta t_\perp$. The effective
contact amplitude of level $q$ is its endpoint weight
\begin{equation}
\Gamma_q=\Gamma|\psi_q(1)|
=
\Gamma\sqrt{\frac{2}{L_M+1}}\sin k_q.
\label{eq:Gamma-q}
\end{equation}

\paragraph{Coherent propagation through a finite normal segment.}
To evaluate propagation through the normal segment, we return to the
original gauge in which the rung hopping is real and the
superconducting phases are carried by the lead surface Green
functions. This representation is unitarily equivalent to the
real-pairing gauge used above and diagonalizes the normal-region rung
hopping in the bonding/antibonding basis. For arbitrary finite $L_M$,
write $N=L_M$. The scalar end-to-end propagator of the uniform serial
segment admits the equivalent representations
\begin{equation}
g_N(z)
=
\frac{t^{N-1}}{\displaystyle\prod_{q=1}^{N}(z-\xi_q)}
=
\sum_{q=1}^{N}\frac{\psi_q(1)\psi_q(N)}{z-\xi_q},
\label{eq:gN-body}
\end{equation}
which coincides with the corner Green function
$[G_M(z)]_{1N}$ in the longitudinal convention
$G_M(z)=(zI+h_M)^{-1}$ of Appendix~\ref{app:general-phi}
[Eq.~\eqref{eq:general-multimode-green}]. The energy-independent
corner cofactor of the serial normal Hamiltonian gives the product
representation above. This property will be used to establish the
finite-segment sign rules in Sec.~\ref{normallevels}. In the
bonding/antibonding basis the two-leg hole end-to-end propagator is
then the pair of rung-shifted copies
\begin{equation}
G_{M,h}^{LR}(z)
=
\sum_{\eta=\pm}P_\eta\,g_N(z-\eta t_\perp),
\label{eq:GMh-body}
\end{equation}
with $P_\eta=(I+\eta\sigma_x)/2$
[Eq.~\eqref{eq:general-multimode-eh}]. Thus, for any finite $N$,
normal-region propagation is carried by one scalar end-to-end
kernel and the two rung branches. The propagator retains coherent contributions from all longitudinal
modes, whose interference enters the fourth-order Josephson response
at the same order as the individual-mode terms.
The corresponding electron propagator and the phase-dependent
fourth-order closed loop with the anomalous lead surface Green
functions are given in Appendix~\ref{app:general-phi}
[Eq.~\eqref{eq:general-multimode-loop}].

The normal-region propagator remains exact for any superconducting
pairing parameters. The Josephson sign also depends on the
superconducting surface Green functions, including their boundary and
finite-energy contributions away from the Kitaev sweet point.
To identify the contributions that select the equilibrium phase,
we next consider $\phi=\pi$, where an additional symmetry
permits an exact decomposition of the junction spectrum.

\subsection{Exact helicity decomposition at \texorpdfstring{$\phi=\pi$}{phi = pi}}\label{low-energy-picture}
After the gauge rotation that renders the intraleg pairing real,
the inter-leg hopping at $\phi=\pi$ can be written as
\begin{equation}
-it_\perp\left(c_1^\dagger c_2-c_2^\dagger c_1\right)
=
-\frac{it_\perp}{2}\left(\gamma_1^A\gamma_2^A+\gamma_1^B\gamma_2^B\right),
\label{eq:trung-secondquant}
\end{equation}
using $c_\ell=(\gamma_\ell^A+i\gamma_\ell^B)/2$, so that in the ladder
BdG representation
$H_\perp=t_\perp\sigma_y\otimes I_{\rm BdG}$
(the Hermiticity is carried by the explicit $i$). For a general phase,
the coupling is the $\delta=\phi-\pi$ expansion of
Eq.~\eqref{eq:trung-delta}.
Each leg is a complete single-leg S/N/S junction in its own right.
We denote by
$H_{\rm ch}^{(L_M)}(\theta)$ the corresponding BdG Hamiltonian matrix
of the isolated single-leg junction, comprising the two superconducting
leads, the finite normal segment of arbitrary length $L_M$, and the
two interfaces, but with the rung hopping removed. Because the two
legs are identical and the intraleg structure is leg-diagonal, the
full $\phi=\pi$ ladder takes the block form
\begin{equation}
H_{\rm lad}^{(L_M)}
=
I_{\rm leg}\otimes H_{\rm ch}^{(L_M)}(\theta)
+
t_\perp\sigma_y\otimes I.
\label{eq:ladder-block}
\end{equation}
Since the two terms act on different tensor factors, they commute,
allowing an exact decomposition into the two helicity sectors
$\sigma_y=\pm1$. Each particle--hole pair $\pm E_n$ is shifted by the
helicity eigenvalues $\pm t_\perp$. Collecting the resulting positive
BdG energies gives $E_n+t_\perp$ and $|E_n-t_\perp|$, whose joint
ground-state contribution is
\begin{equation}
\mathcal E_n(\theta)
=
-\frac12[E_n+t_\perp+|E_n-t_\perp|]
=
-\max[E_n(\theta),t_\perp].
\label{eq:maxformula}
\end{equation}
This identity is exact for arbitrary finite $L_M$ at $\phi=\pi$: it
follows solely from the helicity decomposition of two identical legs
and does not require the Kitaev sweet point. Explicit forms of
$E_n(\theta)$ enter only when quantitative thresholds or residual
harmonics are evaluated.

Equation~\eqref{eq:maxformula} shows that the combined contribution
of two positive BdG levels to the ground-state energy can become
independent of the Josephson phase even when the individual levels
remain dispersive. If $E_n(\theta)<t_\perp$ for all $\theta$,
the opposite phase dependences of the two levels cancel in their sum,
so $\mathcal E_n(\theta)=-t_\perp$ and this pair contributes no
equilibrium Josephson current. For brevity, we call this cancellation
\emph{clamping}, and refer to the corresponding branch contribution
as clamped. If $E_n(\theta)>t_\perp$ over part or all of the phase
cycle, the joint ground-state contribution retains phase dependence.

\subsection{Selective and collective clamping}\label{sec:clamping}
\paragraph{Off-resonant selective clamping.}
We first consider a single detuned normal level with energy
$\varepsilon\neq0$, coupled to the two boundary Majoranas. This
description applies directly to $L_M=1$, where $\varepsilon=\mu_{\rm M}$,
and to an isolated mode $q$ of a finite segment upon replacing
$(\varepsilon,\Gamma)$ with $(\xi_q,\Gamma_q)$. The complete multilevel
response is treated in Sec.~\ref{normallevels}. The hybridized
end-mode energies follow from the corresponding $4\times4$ BdG,
\begin{equation}
e_\pm^2(\theta)
=
\frac{\varepsilon^2+2\Gamma^2
     \pm\sqrt{\varepsilon^4+4\varepsilon^2\Gamma^2
     +4\Gamma^4\sin^2(\theta/2)}}{2},
\label{eq:end-mode-mu}
\end{equation}
which reduces to $\Gamma\sqrt{1\pm\sin(\theta/2)}$ at resonance.
Since $e_-(\theta)$ is maximal at $\theta=0$, whereas $e_+(\theta)$ is
maximal at $\theta=\pi$, their occupied-state contributions $-e_-$ and
$-e_+$ are respectively $0$-phase- and $\pi$-phase-favoring. The two contributions therefore carry competing phase tendencies.
Their combined isolated-channel spectrum retains a zero-energy branch
crossing at $\theta=\pi$ and a discontinuous equilibrium CPR.
A smooth $\pi$-type response emerges when inter-leg coupling selectively
clamps the $0$-phase-favoring contribution.

\begin{figure*}[!t]
\centering
\includegraphics[width=\textwidth]{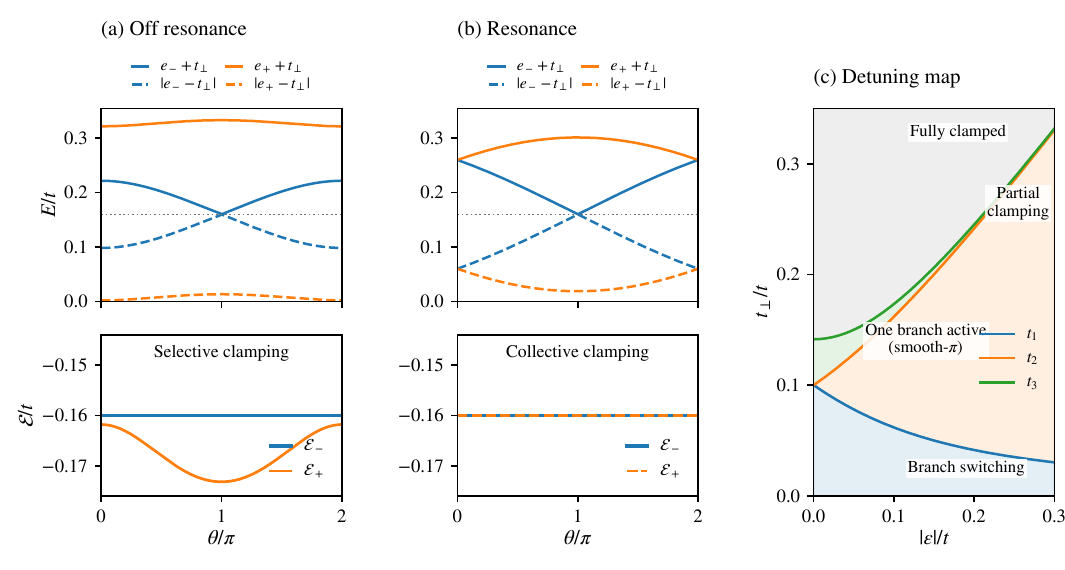}
\caption{Analytical clamping mechanism of the detuning--resonance
transition in the effective isolated-channel representation
(schematic, from the analytic expressions of Eq.~\eqref{eq:end-mode-mu},
with $t=1$ and $\Gamma=0.1t$). The construction applies directly to
$L_M=1$ and to an isolated normal mode $q$ of a finite segment under
$(\varepsilon,\Gamma)\to(\xi_q,\Gamma_q)$.
The upper panels of (a) and (b) plot the helicity-resolved positive
ladder energies $e_\pm+t_\perp$ and $|e_\pm-t_\perp|$. The lower
panels plot the resulting low-energy equilibrium contributions
$\mathcal{E}_{\pm}(\theta)=-\max[e_\pm(\theta),t_\perp]$.
At $\phi=\pi$:
(a) Off resonance ($\varepsilon=0.1t$, $t_\perp=0.16t$, so
$t_{\perp,1}<t_\perp<t_{\perp,2}$): the lower helicity pair mirrors about $t_\perp$,
$(e_-+t_\perp)+|e_--t_\perp|=2t_\perp$, giving a clamped
$\mathcal{E}_-=-t_\perp$, while the upper pair is not
mirror-symmetric and leaves $\mathcal{E}_+$ phase dispersive
(selective clamping).
(b) Resonance ($\varepsilon=0$, $t_\perp=0.16t>\sqrt2\,\Gamma$):
both helicity pairs mirror about $t_\perp$, so
$\mathcal{E}_-=\mathcal{E}_+=-t_\perp$ and $J_{1,\rm low}=0$
(collective clamping). Panels (a) and (b) display only the isolated
low-energy channel sector. Thus $J_{1,\rm low}=0$ in (b) does not
imply a vanishing total Josephson current, and the positive residual
coupling from finite-energy quasiparticles is not shown (derived in
Sec.~\ref{normallevels}).
(c) Detuning--clamping map in the $|\varepsilon|$--$t_\perp$ plane
(axis labels $t_i$ stand for $t_{\perp,i}$, $i=1,2,3$):
the smooth-$\pi$ window $t_{\perp,1}<t_\perp<t_{\perp,2}$ collapses at
resonance ($t_{\perp,1}=t_{\perp,2}=\Gamma$), and the low-energy fully clamped region
starts at
$t_{\perp,3}=\sqrt{\varepsilon^2+2\Gamma^2}$ ($t_{\perp,3}=\sqrt2\,\Gamma$ at
resonance). The off-resonant $\pi$ response is of selective-clamping
type. Resonance closes the window.}
\label{fig:clamping-mechanism}
\end{figure*}

Squaring the clamping condition $e_\pm(\theta)=t_\perp$
yields the same equation for both branches,
\begin{equation}
t_\perp^2(\varepsilon^2+2\Gamma^2-t_\perp^2)
=
\Gamma^4\cos^2\frac{\theta}{2}.
\label{eq:clamping-boundary}
\end{equation}
Thus the lower ($-$) and upper ($+$) branches define two clamping
boundaries $t_\perp=e_\pm(\theta)$ whose extrema are
$t_{\perp,1}=\max_\theta e_-(\theta)$, $t_{\perp,2}=\min_\theta
e_+(\theta)$, and $t_{\perp,3}=\max_\theta e_+(\theta)$. The most
important geometric relation is that the two inner scales are
separated by exactly the detuning,
\begin{equation}
t_{\perp,2}-t_{\perp,1}=|\varepsilon|.
\label{eq:threshold-relations}
\end{equation}
The detuning is therefore identically the width of the selective-clamping
interval. The smooth-$\pi$ window $t_{\perp,1}<t_\perp<t_{\perp,2}$ is
a crossing-free interval between the two clamping boundaries: the
lower $0$-phase-favoring branch is clamped for all $\theta$ while the upper
$\pi$-favoring branch remains dispersive, as illustrated in
Fig.~\ref{fig:clamping-mechanism}(a). At resonance ($\varepsilon=0$), the two thresholds merge
($t_{\perp,1}=t_{\perp,2}=\Gamma$), closing the
selective-clamping interval.
The resulting detuning--clamping structure is summarized in
Fig.~\ref{fig:clamping-mechanism}(c).

With the $\phi=\pi$ commutation
($H_\perp=t_\perp\sigma_y\otimes I$), each detuned branch contributes
$-\max(e,t_\perp)$. In the selective-clamping window the low-energy
ground-state energy is therefore
\begin{equation}
E_{\rm GS}^{\rm low}(\theta)
=
-t_\perp-e_+(\theta),
\qquad
t_{\perp,1}<t_\perp<t_{\perp,2}.
\label{eq:selective-EGS}
\end{equation}
In this interval the lower branch contributes the phase-independent
energy $-t_\perp$, while the upper branch retains its $\pi$-favoring
dispersion. For weak interface coupling, $\Gamma\ll|\varepsilon|$, the
remaining dispersive branch yields the first harmonic
\begin{equation}
J_{1,\rm low}
=
-\frac{\Gamma^4}{4|\varepsilon|^3}+O(\Gamma^6)
<0,
\label{eq:second-order-recovery}
\end{equation}
(see Appendix~\ref{app:detuned-clamping}; off resonance, the effective
cotunneling amplitude is of order $\Gamma^2/|\varepsilon|$, while the
clamped equilibrium harmonic is of order $\Gamma^4$). The ladder $\pi$-type harmonic thus arises from selective clamping:
detuning sets the window width, while $t_\perp$ selects which
branches are clamped. The three thresholds also distinguish the mixed,
smooth-$\pi$, partially clamped, and fully clamped regimes shown in
Fig.~\ref{fig:clamping-mechanism}(c). Their explicit expressions are
given in Appendix~\ref{app:detuned-clamping}.

\paragraph{Resonant collective clamping.}

The resonant limit follows directly from the detuning geometry of
the selective-clamping analysis above. Setting $\varepsilon=0$ collapses
the two inner clamping scales,
\begin{equation}
t_{\perp,1}=t_{\perp,2}=\Gamma,
\qquad
t_{\perp,3}=\max_\theta e_+(\theta)=\sqrt2\,\Gamma,
\label{eq:resonant-thresholds}
\end{equation}
so the finite selective-clamping interval disappears. The resonant
isolated-channel branches follow from Eq.~\eqref{eq:end-mode-mu} at
$\varepsilon=0$,
\begin{equation}
e_\pm(\theta)=\Gamma\sqrt{1\pm\sin(\theta/2)}.
\label{eq:resonant-branches}
\end{equation}
For $t_\perp>\sqrt2\,\Gamma$, both branches satisfy
$e_\pm(\theta)<t_\perp$ for all $\theta$. Thus, by the exact clamping
identity of Eq.~\eqref{eq:maxformula}, their ground-state contributions
coincide,
\begin{equation*}
\mathcal E_+=\mathcal E_-=-t_\perp,
\qquad
E_{\rm GS}^{\rm low}=-2t_\perp,
\end{equation*}
and consequently the corresponding first harmonic and current vanish,
$J_{1,\rm low}=0$ and $I_{\rm low}(\theta)=0$.
In \emph{collective clamping}, the entire resonant low-energy
manifold becomes phase independent and ceases to select the
equilibrium Josephson phase [Fig.~\ref{fig:clamping-mechanism}(b)]. For $t_\perp<t_{\perp,3}=\sqrt2\Gamma$,
at least one resonant branch remains phase dependent or undergoes branch
switching, so complete low-energy clamping has not yet occurred.

For an isolated normal mode $q$, the effective low-energy theory
predicts collective clamping above $t_\perp=\sqrt{2}\,\Gamma_q$, so that
$J_{1,\rm low}=0$. For a finite multilevel segment, a sufficient
condition ensuring the corresponding spectral separation is derived in
Appendix~\ref{app:green-function}. Continuum and higher subgap states
can then carry the residual Josephson current.\cite{Krichevsky2000}
Exact low-energy phase cancellation makes those contributions decide
the coupling sign. That sign is addressed for arbitrary finite serial
segments in Sec.~\ref{normallevels} and realized in closed form for the
one-site junction below.

\subsection{Fourth-order sign rules for finite normal segments}\label{normallevels}
The exact clamping identity determines which quasiparticle contributions
are removed from the equilibrium Josephson energy, but the sign of the
remaining harmonic requires a calculation of the unclamped spectrum.
At the superconducting sweet point this can be achieved for any fixed
finite normal-segment length using a common fourth-order spectral
determinant. In the following, $J_{1,\rm off}^{(4)}$ and
$J_{1,\rm on}^{(4)}$ denote the fourth-order contributions to the first
Josephson harmonic in the specified off- and on-resonant regimes.
The superscript $(4)$ includes the overall factor $\Gamma^4$.

\paragraph{Off and on resonance from one spectral determinant.}
Here the superconducting sweet point means $\mu_S=0$ and
$|\Delta|=t$. The normal chemical potential $\mu_{\rm M}$, and thus
$\{\xi_q\}$, remains free and selects off- versus on-resonance.
Let $P_0(E)$ denote the uncoupled characteristic polynomial at
$\Gamma=0$, after removing the phase-independent outer-Majorana
factor. We separate it into a low-energy spectral factor and a
complementary finite-energy factor $Q(E)$. Off and on resonance these
factors take the forms
\begin{equation}
\begin{aligned}
P_{0,\rm off}(E)&=E^2Q_{\rm off}(E),\\
P_{0,\rm on}(E)&=E^4Q_q(E)
\end{aligned}
\label{eq:P0-factors}
\end{equation}
(Appendix~\ref{app:green-function}). At resonance an additional
normal level enters the low-energy subspace of this uncoupled
reference spectrum, increasing the order of the zero-energy spectral
factor from two to four. Consequently, separating the clamped
low-energy sector requires one additional Taylor subtraction in the
resonant case. The two signs are then fixed by different inequalities
of the resulting frequency kernels.

At the Kitaev sweet point, a boundary electron couples both to the
Majorana zero mode and to the adjacent finite-energy Majorana dimer
at energy $2t$. Retaining both contributions in the superconducting
boundary Green function produces the frequency-dependent factor in
the fourth-order kernel. Off resonance every $\xi_p\neq0$, so only a
constant Taylor term is subtracted from the $E^2$ factor, and the
resulting contribution to the first Josephson harmonic is
\begin{equation}
\begin{aligned}
J_{1,\rm off}^{(4)}
&=
\frac{\Gamma^4}{2\pi}\int_0^\infty\frac{d\omega}{\omega^2}
\Bigg[
\left(\frac{4t^2}{\omega^2+4t^2}\right)^2
|g_N(i\omega)|^2\\
&\qquad\qquad\qquad\qquad
-
|g_N(0)|^2
\Bigg],
\end{aligned}
\label{eq:J1-GF}
\end{equation}
in terms of the end-to-end propagator $g_N$ of
Sec.~\ref{sec:general-spectral}. The propagator itself is strictly
monotone on the Matsubara axis,
\begin{equation}
|g_N(i\omega)|^2
=
\frac{t^{2(N-1)}}{\displaystyle\prod_q(\omega^2+\xi_q^2)}
<
|g_N(0)|^2,
\qquad \omega>0.
\label{eq:gN-monotone}
\end{equation}
Together with the finite-energy factor, this inequality makes the
integrand of Eq.~\eqref{eq:J1-GF} strictly negative, so
$J_{1,\rm off}^{(4)}<0$ for any finite serial $L_M$ in the smooth
single-clamped window. The end-mode projection drops the factor
$(4t^2/(\omega^2+4t^2))^2$: it already yields the correct sign but
underestimates the magnitude. Near an isolated level the Laurent piece
$-\Gamma^4|\psi_q(1)|^4/(4|\xi_q|^3)$ is recovered, with an additional
strictly negative finite-energy correction once the $2t$ poles are
kept (Appendix~\ref{app:green-function}).

At an isolated resonance $\xi_q=0$ the same determinant has the
$E^4$ factor of Eq.~\eqref{eq:P0-factors}. In the resonant clamping
regime
\begin{equation}
\sqrt2\,\Gamma_q<t_\perp<\Lambda_q,
\qquad
\Lambda_q=\min\!\left(2t,\min_{p\neq q}|\xi_p|\right),
\label{eq:res-window}
\end{equation}
this interval is the isolated-mode estimate for clamping the resonant
low pair ($J_{1,\rm low}=0$) while leaving higher states unclamped.
For finite $\Gamma$, a rigorous sufficient interval is
Eq.~\eqref{eq:weyl-on}. The lower edge in Eq.~\eqref{eq:res-window}
is not an exact multilevel threshold. Subtracting the constant and
$E^2$ Taylor terms assigned to the low factor leaves the regular
kernel
\begin{equation}
R_q(\omega)
=
\frac{|\psi_q(1)|^4}{2}
\left(\frac{4t^2}{\omega^2+4t^2}\right)^2
\prod_{p\neq q}
\frac{\xi_p^2}{\omega^2+\xi_p^2},
\label{eq:Rq-main}
\end{equation}
which is the resonant-pole regularization of the same frequency
weight that appears in Eq.~\eqref{eq:J1-GF},
$R_q(\omega)=\frac{\omega^2}{2}(4t^2/(\omega^2+4t^2))^2
|g_N(i\omega)|^2$ at $\xi_q=0$. The residual harmonic
(Appendix~\ref{app:res-theorem})
\begin{equation}
J_{1,\rm on}^{(4)}
=
\frac{\Gamma^4}{\pi}
\int_0^\infty
\frac{
R_q(\omega)-R_q(0)-\frac{\omega^2}{2}R_q''(0)
}{\omega^4}\,d\omega>0
\label{eq:J1-res}
\end{equation}
is positive because $R_q(\omega)=R_q(0)\prod_\lambda(1+\omega^2/a_\lambda)^{-1}$
with $a_\lambda>0$ lies strictly above its $\omega^2$-tangent
(convexity on squared Matsubara frequency). The upper bound
$t_\perp<\Lambda_q$ isolates this residual from the large-$t_\perp$
change of the spectrum. The one-site limit is examined below.

Thus the off- and on-resonant signs are fixed by the monotonicity and
convexity of two regular contributions obtained from the same
sweet-point determinant after the low-energy factors in
Eq.~\eqref{eq:P0-factors} have been removed.

\paragraph{Solvable one-site realization.}
At the Kitaev sweet point $\mu_S=\mu_{\rm M}=0$, $t=|\Delta|$, the
$L_M=1$ junction provides a closed-form realization of the general
sign structure, based on the Kitaev-chain construction in Ref. \citen{A.Yu.Kitaev2001} and the single-chain S/N/S analysis of Ref. \citen{PhysRevB.104.205431}.
Here $R_q(\omega)=\frac12[4t^2/(\omega^2+4t^2)]^2$.
Evaluating Eq.~\eqref{eq:J1-res} gives
\begin{equation}
J_1^{(4)}=+\frac{5}{64}\frac{\Gamma^4}{t^3},
\label{eq:J1-eq37}
\end{equation}
as shown in Appendix~\ref{app:green-function}.
This coefficient is independent of $t_\perp$ in the collectively
clamped window; Eq.~\eqref{eq:weyl-on} gives a sufficient interval
at finite $\Gamma$. The resonant low-energy contribution vanishes,
so the positive residual phase dependence comes from finite-energy
states. Away from $\phi=\pi$, helicity mixing introduces a distinct
low-energy contribution, whose fourth-order coupling is examined next.

\subsection{Resonant Josephson coupling at general inter-leg phase}\label{robustness}
The preceding sign rules follow from the exact helicity decomposition
at $\phi=\pi$. At a general inter-leg phase, helicity mixing modifies
the low-energy response. We now show that, near an isolated
normal-level resonance, this mixing produces a positive leading
fourth-order harmonic, even though its spectral origin differs from
the finite-energy contribution that determines the residual coupling
at $\phi=\pi$.

\paragraph{Weak-inter-leg limit at an isolated resonance.}
In the controlled limit $t_\perp\ll t$ the general-$\phi$ low-energy
sector can be solved in closed form (eight-Majorana projection,
Appendix~\ref{app:general-phi}). For the one-rung realization, along
the leading resonance
$\mu_{\rm M}=t_\perp[\sin(\phi/2)-1]$, the zero-energy-crossing-free condition is
\begin{equation}
t_\perp
>
\Gamma\sqrt{\frac{2}{s_\phi(2-s_\phi)}},
\qquad
s_\phi=\sin\frac{\phi}{2},
\label{eq:rc-phi}
\end{equation}
(reducing to $t_\perp>\sqrt2\,\Gamma$ at $\phi=\pi$), and the
low-energy fourth-order contribution to the first Josephson harmonic is
\begin{equation}
J_{1,\rm low}^{(4)}(\phi)
=
\frac{\Gamma^4}{64t_\perp^3}
\frac{(1-s_\phi^2)(1+3s_\phi)}{s_\phi^3}
>
0
\label{eq:J1-low-phi}
\end{equation}
for $0<\phi<\pi$.
In the weak-inter-leg regime $t_\perp\ll t$, with
$\Gamma\ll t_\perp s_\phi$, the resonant low-energy contribution is
strictly positive for fixed $0<\phi<\pi$. Helicity mixing therefore
restores a positive Josephson harmonic away from the exact clamping
point, with a spectral origin distinct from the finite-energy
contribution at $\phi=\pi$ (at $\phi=\pi/3$ exactly
$15/64\,\Gamma^4/t_\perp^3$).

For an isolated mode $q$ of any fixed finite uniform normal segment,
the projected result follows by replacing $\Gamma$ with $\Gamma_q$
and $\mu_{\rm M}$ with $\xi_q$. The leading alignment is
$\xi_q=t_\perp(s_\phi-1)$, or
$\mu_{\rm M}=-2t\cos k_q+t_\perp(s_\phi-1)$.
At the superconducting sweet point, the controlled
isolated-resonance asymptotics of
Appendix~\ref{app:general-phi}
[Eq.~\eqref{eq:general-resonance-asymptotics}]
bound the remaining spectral contributions under the two
independent small parameters
\begin{equation}
\frac{\Gamma_q}{t_\perp\sin(\phi/2)}\ll1,
\qquad
\frac{t_\perp}{\Lambda_q}\ll1.
\label{eq:two-small-params}
\end{equation}
The first condition controls the interface expansion within the
resonant low-energy sector, whereas the second controls its
separation from the remaining normal and superconducting states.
For each fixed $0<\phi<\pi$ the fourth-order coefficient is then
positive for sufficiently small $t_\perp/\Lambda_q$, with constants
that need not remain uniform as $\phi\to0$ or $\phi\to\pi$.
In this crossing-free limit the leading CPR is sinusoidal,
$I(\theta)=(2e/\hbar)J_{1,\rm on}^{(4)}(\phi;q)\sin\theta+O(\Gamma^6)$.
At $\phi=\pi$ the same low-energy term vanishes and the positive
finite-energy remainder of Eq.~\eqref{eq:J1-res} selects the phase.

\paragraph{Symmetry and local stability about $\phi=\pi$.}
With $\delta=\phi-\pi$, leg-exchange symmetry
($\phi\to-\phi\equiv2\pi-\phi$) makes leg-exchange-invariant
equilibrium observables even in $\delta$, so in a regular
zero-crossing-free sector
\begin{equation}
J_1(\pi+\delta)=J_1(\pi-\delta)
\quad\Longrightarrow\quad
J_1(\pi+\delta)=J_1(\pi)+O(\delta^2).
\label{eq:symmetry-local}
\end{equation}
Since $J_1(\pi)>0$
within the finite-segment resonant clamping regime
[Eq.~\eqref{eq:J1-res}], the resonant $0$-type response is
analytically stable in a finite neighborhood of $\phi=\pi$
(Eq.~\eqref{eq:J1-eq37} provides its closed-form $L_M=1$ realization). The symmetry establishes local stability around $\phi=\pi$;
the wider interval in the representative finite segment is
characterized by the numerical results below.
The detailed $\Gamma$-order closed-loop structure and the
helicity-mixing selection rule are given in
Appendix~\ref{app:general-phi}.

\paragraph{Finite-segment numerical results.}
For the representative parameters $L_M=5$, $\Delta=0.6t$, and
$t_\perp=\Gamma=0.1t$, Fig.~\ref{fig:resonance-diag}(c) shows that
the resonant $0$-type windows extend from $\phi=\pi$ to at least
$\phi=\pi/3$. Additional calculations at $\phi=2\pi/3,\pi/2,\pi/3$
yield an approximately quartic dependence of the first harmonic on
$\Gamma$. These numerical results extend the observed
resonance-induced suppression beyond the sweet-point and weak-coupling
regimes used in the analytical calculations.

\section{Conclusion}

Discrete-level resonance in a finite normal segment changes which quasiparticle contributions determine the equilibrium Josephson coupling of a Kitaev-ladder junction. Off resonance, inter-leg hybridization can selectively clamp the $0$-favoring low-energy ground-state contribution, leaving a dispersive $\pi$-favoring contribution and producing a smooth negative first harmonic within the corresponding parameter window. This mechanism can operate even when the superconducting ladder is bulk-topologically trivial.

At $\phi=\pi$, the exact collective-clamping identity makes the \emph{ground-state energy contributions} of both resonant low-energy branches phase independent above the clamping threshold, although individual quasiparticle levels may remain dispersive. Finite-energy states then select the residual coupling, whose leading first harmonic is positive in the controlled sweet-point regime. At fixed $0<\phi<\pi$, a separate weak-coupling expansion yields a positive resonant low-energy harmonic through helicity mixing. Together, the exact cancellation identity, fourth-order sign rules, and finite-segment calculations show how resonance redistributes the spectral contributions underlying the Josephson $0$- and $\pi$-type responses.

\section*{Acknowledgments}
\begin{acknowledgments}
This work was supported by JST SPRING, Grant No. JPMJSP2114, and the
GP-Spin Program. We thank Shuhei Fukuda for helpful assistance with
the numerical calculations, and Dapeng Yao and Shunji Tsuchiya for
useful discussions related to this work.

We acknowledge the use of the BCLS research system for assistance with computational code development, cross-checking analytical derivations, and manuscript preparation. The authors take full responsibility for the scientific content of this work.
\end{acknowledgments}

\section*{Data Availability}
Selected calculation and figure-generation code, numerical scan
caches, and the figure files used in this manuscript are publicly
available in the GitHub release
\texttt{submission-2026-10-10} at
\url{https://github.com/Slabestone/kitaev-ladder-josephson-resonance/releases/tag/submission-2026-10-10}.\cite{Xie2026Code}
The repository documents the available reproduction workflows and
their scope. The supplied scan caches and figure exports do not
constitute a complete archive of all original simulation outputs.

\appendix
\makeatletter
\renewcommand{\theHequation}{\Alph{section}.\arabic{equation}}
\renewcommand{\theHfigure}{\Alph{section}.\arabic{figure}}
\renewcommand{\theHtable}{\Alph{section}.\arabic{table}}
\makeatother

\section{Recursive Green-Function Implementation}
\label{app:rgf}
Here we collect the recursive Green-function construction used to evaluate
the Josephson current of Sec.~\ref{Model and Formalism}. We use the
four-component Nambu slice basis
$\Psi_x=(c_{1x},c_{2x},c_{1x}^{\dagger},c_{2x}^{\dagger})^T$
(leg-first, particle/hole-grouped ordering). In this basis the
superconducting-lead Hamiltonian takes the block-tridiagonal form
\begin{equation}
\check{H}^{(S,\alpha)}_{xy}
=
\check{u}_S\delta_{xy}
+
\check{t}_{S,\alpha}\delta_{x,y-1}
+
\check{t}_{S,\alpha}^{\dagger}\delta_{x,y+1},
\end{equation}
with the $4\times4$ onsite and hopping blocks
\begin{align}
\check{u}_S &=
\begin{bmatrix}
-\mu_S & -t_\perp & 0 & 0 \\
-t_\perp & -\mu_S & 0 & 0 \\
0 & 0 & \mu_S & t_\perp \\
0 & 0 & t_\perp & \mu_S
\end{bmatrix},
\\
\check{t}_{S,\alpha} &=
\begin{bmatrix}
-t & 0 & \Delta_{1\alpha} & 0 \\
0 & -t & 0 & \Delta_{2\alpha} \\
-\Delta_{1\alpha}^* & 0 & t & 0 \\
0 & -\Delta_{2\alpha}^* & 0 & t
\end{bmatrix}.
\end{align}
The subscript $\mathrm{S}$ labels superconducting slices and
$\alpha=L,R$ the lead. The pairing entries are those of
Sec.~\ref{Model and Formalism},
$\Delta_{\beta\alpha}=|\Delta|e^{-i(\theta_\alpha+\phi_\beta)}$. The
bulk topological classification of the isolated Kitaev ladder
described by Eq.~\eqref{eq:kitaev ladder} is summarized in
Appendix~\ref{Majorana zero modes on the Kitaev ladder}. In the
following analysis, this classification is used only as a reference
for identifying the topological phase boundaries of the
superconducting leads, while the main focus is the junction response
controlled by tunneling geometry and the resonance structure of the
finite normal segment.

The local Green's function in a finite system is defined by first inverting the full matrix and then taking the $(x,y)$ block,
\begin{align}
\check{G}_{xy}^{(n)}(z)
=
\left[
\left(zI-\check{H}^{(n)}\right)^{-1}
\right]_{xy},
\end{align}
with $\check{z}\equiv zI_4$ in the four-component Nambu slice used below. The complex frequency $z$ takes the form:
\begin{equation}
\label{eq6}
z = \begin{cases}
E - i0^+ & \text{(advanced Green's function)} \\
E + i0^+ & \text{(retarded Green's function)} \\
i\omega_m & \text{(Matsubara Green's function)},
\end{cases}
\end{equation}
where $0^+$ denotes a positive infinitesimal.

The local Green's functions are evaluated with the recursive Green's function method.\cite{Sancho_1985, Yada2014, PhysRevB.55.5266, PhysRevB.87.024515, PhysRevB.101.024509} The surface Green functions of the two semi-infinite superconducting leads obey orientation-dependent recursion relations. With the hopping convention of Eq.~\eqref{eq:kitaev ladder}, they are
\begin{align}
\check{g}_{L}^{(n)}
&=
\left[
\check{z}-\check{u}_S
-\check{t}_{S,L}^{\dagger}
\check{g}_{L}^{(n-1)}
\check{t}_{S,L}
\right]^{-1},
\label{eq:lead-recursion-L}
\\
\check{g}_{R}^{(n)}
&=
\left[
\check{z}-\check{u}_S
-\check{t}_{S,R}
\check{g}_{R}^{(n-1)}
\check{t}_{S,R}^{\dagger}
\right]^{-1},
\label{eq:lead-recursion-R}
\end{align}
where the order of $\check{t}_{S,\alpha}$ and $\check{t}_{S,\alpha}^{\dagger}$ follows from the growth direction: the left surface faces its rightmost site and grows rightward ($\check{H}_{n,n-1}=\check{t}^\dagger$), while the right surface faces its leftmost site and grows leftward ($\check{H}_{n,n+1}=\check{t}$). Since the superconducting hopping block is not Hermitian ($\check{t}_{S,\alpha}^{\dagger}\neq\check{t}_{S,\alpha}$), the two recursions are distinct. Here $\check{u}_S$ is the common onsite slice and $\check{t}_{S,\alpha}$ carries the lead phase factors of Eq.~\eqref{eq:kitaev ladder}. The semi-infinite surface Green functions are $\check{g}_\alpha^{S}=\lim_{n\to\infty}\check{g}_{\alpha}^{(n)}$, $\alpha=L,R$.

The finite normal segment is then attached recursively. The normal slice is described by
\begin{align}
\check{u}_M
&=
\begin{bmatrix}
-\mu_{\rm M} & -t_\perp & 0 & 0 \\
-t_\perp & -\mu_{\rm M} & 0 & 0 \\
0 & 0 & \mu_{\rm M} & t_\perp \\
0 & 0 & t_\perp & \mu_{\rm M}
\end{bmatrix},
\\
\check{t}_M
&=
\begin{bmatrix}
-t & 0 & 0 & 0 \\
0 & -t & 0 & 0 \\
0 & 0 & t & 0 \\
0 & 0 & 0 & t
\end{bmatrix},
\label{eq:normal-slice}
\end{align}
obtained from the superconducting slice by replacing $\mu_S\to\mu_{\rm M}$ and setting the pairing amplitude to zero ($\Delta=0$). The interface hopping between the lead and the first normal slice is
\begin{align}
\check{t}_\Gamma
=
\begin{bmatrix}
-\Gamma & 0 & 0 & 0 \\
0 & -\Gamma & 0 & 0 \\
0 & 0 & \Gamma & 0 \\
0 & 0 & 0 & \Gamma
\end{bmatrix}.
\label{eq:interface-hopping}
\end{align}
The finite normal segment is then attached recursively, with the same orientation convention, using the normal-slice blocks of Eq.~\eqref{eq:normal-slice} and the interface hopping of Eq.~\eqref{eq:interface-hopping}. The first normal slice connected to the left superconducting lead is
\begin{align}
\check{\mathcal G}_{L,1}
=
\left[
\check{z}-\check{u}_M
-\check{t}_{\Gamma}^{\dagger}
\check{g}_L^{S}
\check{t}_{\Gamma}
\right]^{-1},
\label{eq:first-slice}
\end{align}
and subsequent normal slices are added rightward through
\begin{align}
\check{\mathcal G}_{L,j}
=
\left[
\check{z}-\check{u}_M
-\check{t}_M^{\dagger}
\check{\mathcal G}_{L,j-1}
\check{t}_M
\right]^{-1},
\qquad
j=2,\ldots,L_M,
\label{eq:normal-recursion}
\end{align}
Starting from the right superconducting lead, the recursion proceeds leftward: the last normal slice adjacent to the right lead is
\begin{align}
\check{\mathcal G}_{R,L_M}
=
\left[
\check{z}-\check{u}_M
-\check{t}_{\Gamma}
\check{g}_R^{S}
\check{t}_{\Gamma}^{\dagger}
\right]^{-1},
\label{eq:last-slice}
\end{align}
and subsequent normal slices are added leftward through
\begin{align}
\check{\mathcal G}_{R,j}
&=
\left[
\check{z}-\check{u}_M
-\check{t}_M
\check{\mathcal G}_{R,j+1}
\check{t}_M^{\dagger}
\right]^{-1},
\nonumber\\
&\qquad j=L_M-1,\ldots,1.
\label{eq:normal-recursion-R}
\end{align}
Here $\check{g}_\alpha^{S}$ denotes the converged surface Green
function obtained from Eqs.~\eqref{eq:lead-recursion-L} and
\eqref{eq:lead-recursion-R}. Although $\check{t}_M=\check{t}_M^{\dagger}$ makes the normal recursions numerically identical, the orientation is kept explicit for consistency with the superconducting case.

Cutting the normal region between sites $i$ and $i+1$, the dressed
half-Green functions $\check{\mathcal G}_{L,i}$ (left lead plus normal
sites $M_1,\ldots,M_i$) and $\check{\mathcal G}_{R,i+1}$ (normal sites
$M_{i+1},\ldots,M_{L_M}$ plus right lead) are connected by the Dyson
equation
\begin{align}
\check{G}_{ii} &= \left[ (\check{\mathcal G}_{L,i})^{-1} - \check{t}_M \check{\mathcal G}_{R,i+1} \check{t}_M^\dagger \right]^{-1}, \\
\check{G}_{i,i+1} &= \check{G}_{ii} \check{t}_M \check{\mathcal G}_{R,i+1},
\end{align}
with the analogous expressions for $\check{G}_{i+1,i+1}$ and
$\check{G}_{i+1,i}$. The Josephson current across the normal-metal link is evaluated using
the expression in Ref. \citen{PhysRevB.63.052512}:
\begin{equation}
I(i) = -\frac{ieT}{\hbar} \sum_{\omega_m} t \, \text{Tr} \left[ \check{G}_{i,i+1} - \check{G}_{i+1,i} \right],
\end{equation}
where the trace is over the electron (Nambu particle) block of the
local Green functions.

\section{Topological Classification of the Kitaev Ladder}
\label{Majorana zero modes on the Kitaev ladder}

For a gapped isolated ladder, the symmetry class determines the
bulk invariant.\cite{PhysRevB.89.174514} At $\phi=0$ or $\pi$ the
Hamiltonian has chiral symmetry and belongs to class BDI, with an
integer winding number $w$.\cite{Ryu2002,Schnyder2008,Kitaev2009,Tanaka2012,Sato2011}
For $0<\phi<\pi$ chiral symmetry is broken and class D has a
$\mathbb Z_2$ invariant $\nu$.\cite{Altland1997} This invariant fixes
the parity of protected boundary Majoranas, rather than the total
number of zero modes in special decoupled limits. The phase boundaries
follow from bulk gap closing; boundary zero modes can also be
constructed directly in the Majorana representation.\cite{Kitaev2009,Sato2011,PhysRevB.89.174514,Maiellaro2018,Kanehira2023JosephsonEB}

At $t_\perp=0$, two independent topological Kitaev chains support two
Majorana pairs, although their combined class-D parity is even.
At $\phi=\pi$ and nonzero $t_\perp$, the exact helicity decomposition
of Sec.~\ref{low-energy-picture} shifts the single-chain end modes
away from zero energy. The bulk may also become gapless, where a
gapped-phase invariant is undefined. These special limits should not
be read as additional protected zero-mode counts from the generic
class-D diagram.

\begin{figure}[htbp]
\centering
\includegraphics[width=\columnwidth]
{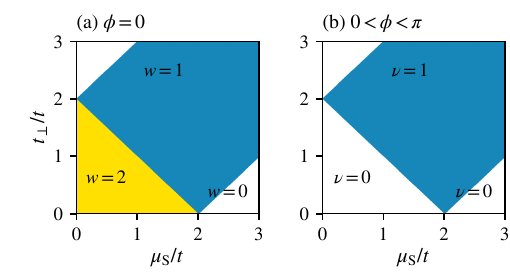}
\caption{Bulk topological phase diagrams of the isolated superconducting
Kitaev ladder as a function of $\mu_S/t$ and $t_\perp/t$ for
(a) $\phi=0$ and (b) $0<\phi<\pi$. The colored regions indicate
(a) the BDI winding number $w$ and (b) the class-D invariant $\nu$.
The gapped trivial regions have $w=0$ or $\nu=0$.
The decoupled limit and $\phi=\pi$ require the separate boundary-mode
qualification given in the text.}
\label{topophase}
\end{figure}

\section{Analytical Derivations}

\subsection{General inter-leg phase \texorpdfstring{$0<\phi<\pi$}{0<phi<pi}}
\label{app:general-phi}

\paragraph{Helicity-mixing selection rule.}
In the gauge where the intrachain pairing is real, the generic
inter-leg coupling decomposes as
\begin{equation}
\begin{aligned}
H_\perp
&=
-t_\perp
\left(
s_\phi\,\sigma_y\tau_0
+
c_\phi\,\sigma_x\tau_z
\right),
\\
s_\phi&=\sin\frac{\phi}{2},\qquad c_\phi=\cos\frac{\phi}{2},
\end{aligned}
\label{eq:helicity-decomp}
\end{equation}
where $\sigma_i$ act in leg space and $\tau_i$ in Nambu space.
Starting from the second-quantized rung hopping in
Eq.~\eqref{eq:trung-delta}, in the ordered leg basis $(c_1,c_2)$, the
BdG matrix is
$H_\perp^{(1,2)}=t_\perp s_\phi\sigma_y\tau_0-t_\perp c_\phi\sigma_x\tau_z$.
The present appendix adopts the exchanged-leg representation obtained by
$U=\sigma_x\tau_0$, which sends $\sigma_y\to-\sigma_y$ and leaves
$\sigma_x$ unchanged, yielding Eq.~\eqref{eq:helicity-decomp}. At
$\phi=\pi$ one has $H_\perp^{(1,2)}=+t_\perp\sigma_y\tau_0$ as in
Sec.~\ref{low-energy-picture}, while $UH_\perp^{(1,2)}U^\dagger=-t_\perp\sigma_y\tau_0$.
The two helicity sectors are merely relabeled and the spectrum is unchanged.
Subsequent Majorana projections in this appendix use the same exchanged-leg
convention. The mixing channel $\sigma_x\tau_z$ anticommutes with the
helicity channel $\sigma_y\tau_0$, so every insertion of
$\sigma_x\tau_z$ reverses the helicity. Consequently,
helicity-preserving closed processes contain an even number of
helicity-mixing insertions, and in an analytic
zero-energy-crossing-free sector diagonal effective corrections are
even in $c_\phi=\cos(\phi/2)$. Because the resonant low-energy
Josephson harmonic vanishes exactly at $\phi=\pi$ (where $c_\phi=0$),
its leading departure from the clamped point necessarily contains the
factor $\cos^2(\phi/2)$.

\paragraph{Exact normal-mode propagation for finite $L_M$.}
The normal segment contains no pairing, so its longitudinal modes can
be eliminated before evaluating the superconducting boundary response.
Here we use the original gauge of Eq.~\eqref{eq:kitaev ladder}, with
real normal-region rung hopping and the lead pairing phases carried by
$\check g_{L,R}^S$. No leg-dependent gauge rotation is made in the
normal segment. Write $N=L_M$, $r=t_\perp$, and let $h_M$ denote the real, uniform
single-leg electron block with onsite entries $-\mu_{\rm M}$ and
nearest-neighbor hoppings $-t$, as in Sec.~\ref{normallevels}.
All subsequent normal-region formulas are built from one scalar end-to-end
propagator,
\begin{equation}
\begin{aligned}
g_N(z)
&=\big[(zI+h_M)^{-1}\big]_{1N}
=\frac{1}{t\,U_N\!\left[(z-\mu_{\rm M})/(2t)\right]}\\
&=\sum_{q=1}^{N}\frac{\psi_q(1)\psi_q(N)}{z-\xi_q},
\end{aligned}
\label{eq:general-multimode-green}
\end{equation}
where $U_N$ is the Chebyshev polynomial of the second kind and
$\xi_q=\mu_{\rm M}+2t\cos[\pi q/(N+1)]$. The Chebyshev identity is the
corner cofactor of the tridiagonal normal Hamiltonian. The modal sum
uses the same poles as Sec.~\ref{normallevels}. In the eigenbasis of
$\sigma_x$, let $P_\eta=(I+\eta\sigma_x)/2$, $\eta=\pm$. The exact
electron and hole end-to-end propagators of the two-leg normal region
are then
\begin{equation}
\begin{aligned}
G_{M,h}^{LR}(z)
&=\sum_{\eta=\pm}P_\eta\,g_N(z-\eta r),\\
G_{M,e}^{LR}(z)
&=-\sum_{\eta=\pm}P_\eta\,g_N(-z-\eta r).
\end{aligned}
\label{eq:general-multimode-eh}
\end{equation}
These identities are exact for any finite $N$, independent of the
sweet point and of any single-mode truncation. The formulas here use
the uniform serial segment specified above. The corner-cofactor
representation in Appendix~\ref{app:green-function} also applies to
real nonuniform serial hoppings. The signs follow from
the physical normal BdG blocks
$h_M\otimes I-rI\otimes\sigma_x$ and
$-h_M\otimes I+rI\otimes\sigma_x$, respectively. In particular, the
hole propagator has poles at $\xi_q+\eta r$, consistently with
Eq.~\eqref{noraml energy}. Since the normal Hamiltonian is real and
symmetric, $G_{M,e}^{LR}(i\omega)=-[G_{M,h}^{LR}(i\omega)]^*$,
where the conjugation acts elementwise in leg space. At
$\mu_{\rm M}=0$, the same Chebyshev identity gives
$g_N(-z)=(-1)^N g_N(z)$. Odd $N$ also has the longitudinal level
$\xi_{(N+1)/2}=0$, whereas even $N$ is detuned from zero. These
facts distinguish the two geometries used in the numerical examples,
but parity alone does not fix the Josephson sign: the full coherent
propagator products and superconducting surface response still
enter the frequency integral.

For clarity, let $\check G_M^{ab}(z)$ denote the corresponding full
leg--Nambu normal-region propagator between contacts $a,b=L,R$,
let $\check g_\alpha^S(z)$ be the lead surface Green function with the
appropriate contact orientation, and let $\tau_z$ act in Nambu space.
The equilibrium Josephson response can be expressed through a
phase-dependent Green-function determinant, as in established treatments
of multilevel superconducting junctions.\cite{DellAnna2007} Expanding
this determinant to fourth order in the interface hopping yields the
phase-dependent closed propagation process connecting the two
superconducting leads. Let $D(z)$ denote the relative spectral
determinant obtained after integrating out the superconducting leads.
Upon integrating out the two leads, the cross-contact part of the
fourth-order relative spectral determinant is
\begin{equation}
\begin{aligned}
\left[\delta_\theta\ln D(z)\right]_{\Gamma^4}
={}&-\Gamma^4\,\delta_\theta\operatorname{tr}\bigl[
\check G_M^{RL}(z)\tau_z\check g_L^S(z)\tau_z\\
&\qquad\times\check G_M^{LR}(z)\tau_z
\check g_R^S(z)\tau_z\bigr].
\end{aligned}
\label{eq:general-multimode-loop}
\end{equation}
Here $\delta_\theta$ extracts the nonconstant phase-dependent part of
an expression, rather than its difference from the value at
$\theta=0$, consistently with Appendix~\ref{app:green-function}.
Phase-independent single-contact loops have been omitted. For the
ground-state energy difference, this term is evaluated at $\theta$ and
at the reference phase $0$ and subtracted. The trace is
over the four contact leg--Nambu components. In the full trace of
Eq.~\eqref{eq:general-fourth-kernel}, the phase-dependent loop occurs
with four cyclic starting points. This factor cancels the $1/4$ in
the fourth-order logarithm expansion and fixes the normalization of
Eq.~\eqref{eq:general-multimode-loop}. The ground-state energy and
its first harmonic then follow from the spectral integral and
Eq.~\eqref{eq:jn-def}, using the same BdG counting convention. Equation
\eqref{eq:general-multimode-loop} is an identity for the $\Gamma^4$
coefficient when the disconnected problem is gapped. Although the
normal region may contain arbitrarily many modes, the trace involves
only the four leg--Nambu degrees of freedom at each contact. The
mode sum must be performed at the propagator level: products of
Eq.~\eqref{eq:general-multimode-eh} contain $q\ne p$ interference
terms at the same order in $\Gamma$. The strictly detuned
$r\to0^+$ limit also requires care, because the lead Majorana
splitting vanishes in that limit: extracting the $\Gamma^4$
coefficient at fixed $r>0$ precedes any $r\to0^+$ asymptotics.

\paragraph{Low-energy degrees of freedom selected by detuning.}
For a strictly detuned finite normal segment, $\min_q|\xi_q|>0$.
If $r$ is small compared with this detuning scale, its resolvent is
analytic near zero energy and the normal modes can be eliminated
from the low-energy description. The retained junction subspace
then consists of four Majoranas, two at each superconducting
contact. The resulting boundary self-energy retains the coherent normal-mode
sum and its intermode interference in the Josephson response. At an isolated resonance the selected normal
mode instead lies on the $O(r)$ energy scale of the boundary
splitting. Its two leg fermions must be retained along with the
four junction Majoranas, producing the eight-Majorana subspace
used below. Thus resonance changes the low-energy degrees of
freedom that must be kept explicitly. The full boundary identity
above applies in both cases wherever the disconnected problem is
gapped.

\paragraph{Eight-Majorana low-energy model (weak $t_\perp/t$).}
For $r=t_\perp\ll t$ at the sweet point, each lead contributes two
junction Majoranas $\gamma_{L1},\gamma_{L2},\gamma_{R1},\gamma_{R2}$
and the normal rung two fermions $c_\beta=(a_\beta+ib_\beta)/2$
(eight Majoranas,
$\mathcal H_{\rm low}=\frac{i}{4}\boldsymbol{\gamma}^{T} A\,\boldsymbol{\gamma}$).
In this local representation, boundary Majoranas are indexed by lead first
and leg second, which differs from the global fermion ordering
$(\beta,\alpha,x)=(\mathrm{leg},\mathrm{region},\mathrm{site})$ of
Sec.~\ref{Model and Formalism}. The end-Majorana
splitting is $rs_\phi$, the normal levels are $\mu_{\rm M}\pm r$, and the
leading resonance is $\mu_{\rm M}=r(s_\phi-1)$, $s_\phi=\sin(\phi/2)$. The interface
coupling follows from direct Majorana projection of the electron
hopping, with
$\chi_{L\beta}=-\theta/4+\phi_\beta/2$,
$\chi_{R\beta}=+\theta/4+\phi_\beta/2$:
\begin{equation}
\begin{aligned}
\mathcal H_{\Gamma,L}&=\frac{i\Gamma}{2}\sum_\beta\gamma_{L\beta}
(a_\beta\cos \chi_{L\beta}-b_\beta\sin \chi_{L\beta}),\\
\mathcal H_{\Gamma,R}&=-\frac{i\Gamma}{2}\sum_\beta\gamma_{R\beta}
(a_\beta\sin \chi_{R\beta}+b_\beta\cos \chi_{R\beta}).
\end{aligned}
\label{eq:app-gamma-Majorana}
\end{equation}
With $E_{\rm GS}^{\rm low}=-\frac14\operatorname{Tr}\sqrt{-A^2}$,
zero-energy crossing is $\operatorname{Pf}A=0$, which at resonance
gives
\begin{equation}
\operatorname{Pf}A
=
r^4s_\phi^3(s_\phi-2)+2\Gamma^2r^2s_\phi^2-\Gamma^4\cos^2\frac{\theta}{2},
\label{eq:app-Pf}
\end{equation}
hence the zero-energy-crossing-free condition Eq.~\eqref{eq:rc-phi}.
The Pfaffian determines whether the low-energy spectrum contains a
zero-energy crossing, but the Josephson harmonic depends on the sum of
the positive quasiparticle energies. Let
$\sigma_1=\sum_{a=1}^{4}E_a$, so that
\begin{equation}
E_{\rm GS}^{\rm low}
=
-\frac14\operatorname{Tr}\sqrt{-A^2}
=
-\frac12\sigma_1.
\label{eq:app-EGS-low-sum}
\end{equation}
At $\Gamma=0$ the four positive energies are
$rs_\phi$, $rs_\phi$, $rs_\phi$, and $r(2-s_\phi)$, hence
$\sigma_1=2r(1+s_\phi)$.
To obtain the phase-dependent sum we combine the Pfaffian with the
coefficients of the characteristic polynomial
$\det(xI-A)=x^8+e_1x^6+e_2x^4+e_3x^2+e_4$,
where $e_k$ are the elementary symmetric polynomials in
$\{E_a^2\}$ and $\sigma_4=E_1E_2E_3E_4=-\operatorname{Pf}A$
in the crossing-free sector.
The phase-dependent fourth-order pieces
($\delta_\theta$ extracts the nonconstant $\theta$ dependence) are
\begin{equation}
\begin{aligned}
\relax[\delta_\theta e_2]_{\Gamma^4}
&=
\Gamma^4\cos\theta,
\\
[\delta_\theta e_3]_{\Gamma^4}
&=
-2\Gamma^4r^2(3s_\phi^2+s_\phi-1)\cos\theta,
\\
[\delta_\theta\sigma_4]_{\Gamma^4}
&=
\frac{\Gamma^4}{2}\cos\theta
\end{aligned}
\label{eq:app-spectral-invariants-fourth}
\end{equation}
($[\delta_\theta e_1]_{\Gamma^4}=0$).
Matching these coefficients through the elementary
symmetric-polynomial relations between $\{E_a\}$ and $\{E_a^2\}$,
with the $\Gamma=0$ values
$\sigma_2=6r^2s_\phi$,
$\sigma_3=2r^3s_\phi^2(3-s_\phi)$, and
$\sigma_4=r^4s_\phi^3(2-s_\phi)$
(and with no phase-dependent term at second order), gives
\begin{equation}
[\delta_\theta\sigma_1]_{\Gamma^4}
=
\frac{\Gamma^4(1-s_\phi^2)(1+3s_\phi)}{32r^3s_\phi^3}
\cos\theta.
\label{eq:app-sigma1-fourth}
\end{equation}
Since $E_{\rm GS}^{\rm low}=-\sigma_1/2$, the low-energy first
Josephson harmonic is
\begin{equation}
J_{1,\rm low}^{(4)}(\phi)
=
\frac{\Gamma^4}{64t_\perp^3}
\frac{(1-s_\phi^2)(1+3s_\phi)}{s_\phi^3},
\label{eq:app-J1-low-phi}
\end{equation}
positive for $0<\phi<\pi$ ($15/64$ at $\phi=\pi/3$). Near
$\phi=\pi+\delta$ it reduces to
$\Gamma^4\delta^2/(64r^3)+O(\delta^4)$. The divergence as
$\phi\to0$ (or $r\to0$) marks the
breakdown of $\Gamma\ll rs_\phi$ (another breakdown of the expansion near the crossing), not a
physical divergence.

Allowing $\xi_q$ to vary independently of the alignment in the same
eight-Majorana Hamiltonian connects its detuned and resonant limits.
Within this isolated-mode projection, helicity mixing away from
$\phi=\pi$ can weaken the negative detuned contribution. For a finite
multilevel junction, the detuned response is determined by the
coherent normal-mode propagator rather than by independent
isolated-mode contributions. In particular, the $L_M=4$,
$\mu_{\rm M}=0$ example in Sec.~\ref{results} contains several
detuned normal modes and cannot be represented by a single selected $q$.

\paragraph{Controlled isolated-resonance asymptotics for finite $L_M$.}
Consider the superconducting sweet point, a fixed finite uniform
normal segment, and an isolated mode $q$. Fix $0<\phi<\pi$ and take
the leading alignment
$\mu_{\rm M}=-2t\cos k_q+r(s_\phi-1)$.
In this paragraph $\Lambda_q$ is the isolation scale of
Eq.~\eqref{eq:res-window} evaluated at $r=0$ on this trajectory.
Two small parameters are kept distinct throughout:
$\Gamma_q/(r s_\phi)\ll1$ and $r/\Lambda_q\ll1$.
First take $\Gamma\to0$ at fixed $r>0$ to define the fourth-order
coefficient, and then $r/\Lambda_q\to0$. The result is not an
expansion along $\Gamma\sim r$, nor a uniform statement as
$\phi\to0$ or $\phi\to\pi$.

To organize the complete fourth-order response of the finite-length
junction and control contributions outside the eight-Majorana subspace,
we expand the relative spectral determinant. This construction
separates the low-energy contribution from its high-energy remainder.
Gauge the lead Josephson phases into the interface hoppings and write
$H(\theta)=H_0+\Gamma V(\theta)$, with
$G_0(i\omega)=(i\omega-H_0)^{-1}$. Here $H_0$ describes the
disconnected regions and is gapped for fixed $0<\phi<\pi$ and
sufficiently small $r/\Lambda_q$. The relative ground-state energy is
\begin{equation}
\begin{aligned}
\Delta E_{\rm GS}(\theta)
={}&
-\frac{1}{2\pi}\int_0^\infty d\omega\,
\operatorname{Re}\ln\det\!\Bigl[\\
&\quad(i\omega-H(\theta))(i\omega-H(0))^{-1}\Bigr].
\end{aligned}
\label{eq:rel-det-EGS}
\end{equation}
Using $\ln\det M=\operatorname{Tr}\ln M$, the determinant can be
expanded in the interface coupling. The first phase-dependent
contribution arises at fourth order in $\Gamma$: a closed process must
visit both superconducting contacts to acquire their relative phase,
so the shortest contributing path is the sequence
$L\to M\to R\to M\to L$. The fourth-order term of
$\operatorname{Tr}\ln(I-\Gamma G_0V)$ then gives
\begin{equation}
\begin{aligned}
\Delta E_{\rm GS}^{(4)}(\theta)
={}&
\frac{\Gamma^4}{8\pi}\int_0^\infty d\omega\,\\
&\times\operatorname{Re}\operatorname{Tr}\Bigl[
(G_0V(\theta))^4-(G_0V(0))^4\Bigr].
\end{aligned}
\label{eq:DeltaEGS-fourth}
\end{equation}
Finally, we extract the first Josephson harmonic with
Eq.~\eqref{eq:jn-def}. The reference-phase term is independent of the
Fourier variable $\theta$ and therefore vanishes under the projection,
yielding
\begin{equation}
\begin{aligned}
J_{1,\rm on}^{(4)}(\phi;q)
={}&-\frac{\Gamma^4}{8\pi^2}
\int_0^{2\pi}d\theta\,\cos\theta\\
&\times\int_0^\infty d\omega\,
\operatorname{Re}\operatorname{Tr}[(G_0V)^4].
\end{aligned}
\label{eq:general-fourth-kernel}
\end{equation}
All integrals here refer to the zero-temperature limit.

To control the modes omitted by the projection, let
$\Pi_{\rm low}$ be the spectral projection of $H_0$ onto the
eight modes arising from the four junction Majoranas and the
resonant normal mode, and $\Pi_{\rm high}=I-\Pi_{\rm low}$.
For semi-infinite leads there are no remote-end Majoranas in this
partition. At small $r/\Lambda_q$, perturbation of this isolated
spectral subspace shifts its projection by $O(r/\Lambda_q)$.
Define $G_{\rm low/high}=\Pi_{\rm low/high}G_0$. The uncoupled
low energies have magnitudes at least $rs_\phi/2$, while the high
energies have magnitudes at least $\Lambda_q/2$, giving
\begin{equation}
\begin{aligned}
\|G_{\rm low}\|&\le(\omega^2+r^2s_\phi^2/4)^{-1/2},\\
\|G_{\rm high}\|&\le(\omega^2+\Lambda_q^2/4)^{-1/2}.
\end{aligned}
\label{eq:low-high-resolvent-bounds}
\end{equation}
Because $V$ has finite rank and bounded norm, these bounds also
control each trace in Eq.~\eqref{eq:general-fourth-kernel}.
As a representative estimate, write $a=rs_\phi/2$ and
$b=\Lambda_q/2$. A contribution with three low-energy and one
high-energy resolvent satisfies
\begin{equation}
\int_0^\infty
\frac{d\omega}
{(\omega^2+a^2)^{3/2}(\omega^2+b^2)^{1/2}}
\le
\frac1{a^2b}.
\label{eq:resolvent-bound-example}
\end{equation}
Contributions with two or more high-energy resolvents are
asymptotically smaller for fixed $s_\phi>0$ and
$r/\Lambda_q\to0$; altogether the high-energy remainder is
$O_{q,\phi}(\Gamma^4/(r^2\Lambda_q))$, with constants independent of
lead length. This controls direct high-energy mixing and does not
replace the low-energy projection error below.

The leading eight-Majorana matrices themselves follow by projection
of the bare disconnected Hamiltonian, rather than by assumption.
On a uniform normal segment the longitudinal modes
$f_{\beta q}=\sum_j\psi_q(j)\,c_{\beta j}$ diagonalize both the
chain hopping and the rung coupling, so finite $r$ does not mix
distinct $q$. Along the resonance trajectory
$\xi_q=r(s_\phi-1)$, the selected-mode block
$\mathcal H_{M,q}$ together with the sweet-point end-Majorana
splitting
$\mathcal H_{\perp,\rm end}=(irs_\phi/2)(\gamma_{L1}\gamma_{L2}
+\gamma_{R1}\gamma_{R2})$
yields the first-order eight-mode BdG matrix $H_{8,0}(r)$ whose
positive energies are
$\{rs_\phi,rs_\phi,rs_\phi,r(2-s_\phi)\}$.
The interfaces attach only at the segment endpoints; with
$\psi_q(L_M)=(-1)^{q+1}\psi_q(1)$ both contacts carry the weight
$\Gamma_q=\Gamma|\psi_q(1)|$, and the common endpoint sign is
absorbed into the right Majoranas without changing $\theta$.
The resulting interface matrix is Eq.~\eqref{eq:app-gamma-Majorana}
with $\Gamma\to\Gamma_q$. Write $H_0(r)=H_0(0)+rW$ at this
alignment. The eight-dimensional zero-energy eigenspace of
$H_0(0)$ is isolated by the gap $\Lambda_q$, so standard
Riesz-projection perturbation estimates apply for sufficiently
small $r/\Lambda_q$. A Riesz contour with the resolvent identity
gives
\begin{equation}
\|\Pi_{\rm low}(r)-\Pi_0\|
\le C_{q,\phi}\,\frac{r}{\Lambda_q},
\label{eq:low-projection-shift}
\end{equation}
where $\Pi_0$ is the $r=0$ projector.
A near-identity unitary identification of the perturbed low
subspace with this fixed eight-dimensional space produces the
BdG operators $H_{\rm eff}$ and $V_{\rm eff}$ satisfying
\begin{equation}
\begin{aligned}
\|H_{\rm eff}-H_{8,0}\|&\le C_{q,\phi}\,r^2/\Lambda_q,\\
\|V_{\rm eff}-V_8\|&\le C_{q,\phi}\,r/\Lambda_q,
\end{aligned}
\label{eq:low-projection-errors}
\end{equation}
Here $V_8$ denotes the interface matrix normalized per unit
$\Gamma$, with endpoint weight
$\Gamma_q/\Gamma=|\psi_q(1)|$.
The physical interface coupling in the projected Hamiltonian is
therefore $\Gamma V_8$, with amplitude $\Gamma_q$.
(Here $H_{8,0}$ is the single-particle BdG matrix associated with
$\mathcal H_{8,0}=\mathcal H_{\perp,\rm end}+\mathcal H_{M,q}$,
not the second-quantized form itself.)
Set $G_8=(i\omega-H_{8,0})^{-1}$ and compare with the identified
low resolvent. The resolvent identity and
Eq.~\eqref{eq:low-high-resolvent-bounds} imply
\begin{equation}
\|G_{\rm eff}-G_8\|
\le\frac{C_{q,\phi}\,r^2}{\Lambda_q(\omega^2+a^2)},
\qquad a=rs_\phi/2.
\label{eq:low-resolvent-error}
\end{equation}
Telescoping the four factors in the trace therefore gives
\begin{equation}
\begin{aligned}
&\left|\operatorname{Tr}(G_{\rm eff}V_{\rm eff})^4
-\operatorname{Tr}(G_8V_8)^4\right|\\
&\quad\le\frac{C_{q,\phi}}{\Lambda_q}
\left[\frac{r}{(\omega^2+a^2)^2}
+\frac{r^2}{(\omega^2+a^2)^{5/2}}\right].
\end{aligned}
\label{eq:low-fourth-trace-error}
\end{equation}
The two integrals are $\pi/(4a^3)$ and $2/(3a^4)$; with the
$\Gamma^4$ prefactor both are
$O_{q,\phi}(\Gamma^4/(r^2\Lambda_q))$.
The low block has dimension eight and the interface operator has
finite rank, so the constants are independent of lead length.
Combining the high-energy remainder with this projection error and
Eq.~\eqref{eq:app-J1-low-phi} yields
\begin{equation}
\begin{aligned}
J_{1,\rm on}^{(4)}(\phi;q)
={}&\frac{\Gamma_q^4}{64r^3}
\frac{(1-s_\phi^2)(1+3s_\phi)}{s_\phi^3}+O_{q,\phi}\!\left(\frac{\Gamma^4}{r^2\Lambda_q}\right).
\end{aligned}
\label{eq:general-resonance-asymptotics}
\end{equation}
The positive leading term determines the fourth-order sign for
sufficiently small $r/\Lambda_q$ at fixed $q$, $L_M$, and
$0<\phi<\pi$, with no uniformity assumed at the phase endpoints.
At $\phi=\pi$ the leading low-energy term vanishes and the separate
finite-energy result of Eq.~\eqref{eq:J1-res} applies.

For semi-infinite leads, $(G_0V)^4$ is finite rank, so the trace is
well defined. The determinant expansion uses the finite-rank
relative determinant $\det(I-\Gamma G_0V)$, rather than an
absolute infinite-dimensional determinant or an extensive ground-state
energy. The phase-independent bulk cancels in the energy difference,
and the bounds above apply to this local interface contribution.

\subsection{Explicit clamping thresholds and weak-coupling expansion}
\label{app:detuned-clamping}

For the detuned isolated-channel model of Sec.~\ref{sec:clamping}, the branch dispersions $e_\pm(\theta)$ are given in Eq.~\eqref{eq:end-mode-mu}. Solving the clamping condition $e_\pm(\theta)=t_\perp$ [Eq.~\eqref{eq:clamping-boundary}] yields the three explicit threshold scales:
\begin{equation}
\begin{aligned}
t_{\perp,1}&=\frac{\sqrt{\varepsilon^2+4\Gamma^2}-|\varepsilon|}{2},
\quad
t_{\perp,2}=\frac{\sqrt{\varepsilon^2+4\Gamma^2}+|\varepsilon|}{2},\\
t_{\perp,3}&=\sqrt{\varepsilon^2+2\Gamma^2},
\end{aligned}
\label{eq:app-three-scales}
\end{equation}
which identically satisfy $t_{\perp,2}-t_{\perp,1}=|\varepsilon|$.

In the smooth-$\pi$ window $t_{\perp,1}<t_\perp<t_{\perp,2}$, the lower branch is clamped ($\mathcal E_-=-t_\perp$), while expanding the dispersive upper branch for $\Gamma\ll|\varepsilon|$ yields
\begin{equation}
e_+(\theta) = |\varepsilon|+\frac{\Gamma^2}{|\varepsilon|}
     -\frac{\Gamma^4}{2|\varepsilon|^3}
     \left[1+\cos^2\frac{\theta}{2}\right]+O(\Gamma^6).
\label{eq:app-eplus}
\end{equation}
Substituting this expansion into $E_{\rm GS}^{\rm low}(\theta)=-t_\perp-e_+(\theta)$ directly recovers the leading negative harmonic $J_{1,\rm low}=-\Gamma^4/(4|\varepsilon|^3)+O(\Gamma^6)$ stated in Eq.~\eqref{eq:second-order-recovery}. The complete fourth-order expression including finite-energy lead quasiparticles is given in Eq.~\eqref{eq:app-J1-off-full}. For $t_\perp>t_{\perp,3}$, both low-energy branches are clamped ($E_{\rm GS}^{\rm low}=-2t_\perp$), leaving the low-energy manifold completely phase independent.

\subsection{Unified fourth-order spectral determinant}
\label{app:green-function}
\label{app:res-theorem}
The off- and on-resonant fourth-order formulas of
Sec.~\ref{normallevels}
[Eqs.~\eqref{eq:J1-GF} and \eqref{eq:J1-res}] are two cases of one
sweet-point characteristic determinant at $\phi=\pi$. Throughout,
$g_N(z)\equiv[G_M(z)]_{1,L_M}$ coincides with Eq.~\eqref{eq:gN-body},
$t_j$ is the longitudinal bond hopping ($t_j=t$ for the uniform
chain), $N\equiv L_M$, and $h_M$ is the physical normal electron block
with eigenvalues $\{-\xi_q\}$. Thus $G_M(z)=(zI+h_M)^{-1}$ as in
Sec.~\ref{sec:general-spectral}. Here $\delta_\theta$ extracts the
nonconstant phase-dependent part of an expression, rather than its
difference from the value at $\theta=0$ (and rather than
$\partial_\theta$). The derivation assumes the Kitaev sweet point
$\mu_S=0$, $|\Delta|=t$, the contact-reduced leads below, weak
interface coupling, and the stated low--high spectral windows. It does
not extend unchanged to general $\phi$ or arbitrary $\Gamma$. The modal
weights below use the uniform chain. The corner-cofactor identity
itself also permits real nonuniform bonds.

\paragraph{Common fourth-order identity.}
At $\phi=\pi$ the ladder decomposes into helicity sectors, so it
suffices to treat a single-leg S/N/S junction and restore the ladder
weight by the clamping identity. The unique $O(\Gamma^4)$ process that
depends on $\theta=\theta_R-\theta_L$ is the closed loop
$L\to M\to R\to M\to L$. The sweet-point anomalous boundary Green functions are
\begin{equation}
g_L^{eh}(E)
=
-\frac{2t^2e^{-i\theta_L}}{E(E^2-4t^2)},
\qquad
g_R^{eh}(E)
=
+\frac{2t^2e^{-i\theta_R}}{E(E^2-4t^2)},
\end{equation}
with opposite contact orientations. Same-contact $he$ components carry
the conjugate phases. At $\mu_S=0$ and $|\Delta|=t$ the Kitaev chain
dimerizes in the Majorana basis,
$\mathcal H_{\rm chain}=it\sum_j\gamma_j^B\gamma_{j+1}^A$. The boundary
electron $c_1=(\gamma_1^A+i\gamma_1^B)/2$ couples only to the zero mode
$\gamma_1^A$ and the adjacent dimer $(\gamma_1^B,\gamma_2^A)$ of energy
$2t$. All farther dimers are dynamically decoupled from the contact.
Hence the boundary Green function of a semi-infinite sweet-point lead
coincides exactly with that of the two-site contact-reduced model used
below, not as a finite-lead approximation, but as a consequence of
Majorana dimerization. Their phase-carrying product is
\begin{equation}
g_L^{eh}g_R^{he}+g_R^{eh}g_L^{he}
=
-\frac{8t^4\cos\theta}{E^2(E^2-4t^2)^2}.
\label{eq:g-cross}
\end{equation}
In the electron--hole Nambu basis the bare normal segment is
$\mathrm{diag}(h_M,-h_M^T)$. Write
$M_e(E)=EI-h_M$,
$M_h(E)=EI+h_M^T$,
and $D_{e,h}=\det M_{e,h}$. The end-to-end cofactors of the serial
chain are energy independent: with electron-block off-diagonals
$+t_j$ and hole-block off-diagonals $-t_j$,
\begin{equation}
\begin{aligned}
C_e^{1N}
&=
D_e\,(M_e^{-1})_{1N}
=
(-1)^{N-1}\prod_{j=1}^{N-1}t_j,
\\
C_h^{N1}
&=
D_h\,(M_h^{-1})_{N1}
=
\prod_{j=1}^{N-1}t_j,
\end{aligned}
\label{eq:corner-C}
\end{equation}
hence
$C_e^{1N}C_h^{N1}=(-1)^{N-1}\prod_jt_j^2$,
independent of $E$ and of $\mu_{\rm M}$. Integrating out the leads by
a Schur complement gives
\begin{equation}
S(E)
=
\begin{pmatrix}
M_e-\Sigma^{ee}&F\\
\bar F&M_h-\Sigma^{hh}
\end{pmatrix},
\end{equation}
where the anomalous blocks $F,\bar F$ are $O(\Gamma^2)$ and localized
at the two contacts
($\Sigma^{ab}=\Gamma^2[g_L^{ab}|1\rangle\langle1|
+g_R^{ab}|N\rangle\langle N|]$ up to the interface
particle--hole sign convention absorbed into $F,\bar F$). Normal
self-energies carry no relative Josephson phase. The block
determinant expands as
\begin{equation}
\det
\begin{pmatrix}
A&F\\
\bar F&B
\end{pmatrix}
=
\det A\,\det B\,
\bigl[
1-\operatorname{Tr}(A^{-1}FB^{-1}\bar F)+\cdots
\bigr],
\end{equation}
so the $O(\Gamma^4)$ phase-dependent piece is
\begin{equation}
\begin{aligned}
\delta_\theta\det S
={}&
-\Gamma^4
\bigl(
g_L^{eh}g_R^{he}
+
g_R^{eh}g_L^{he}
\bigr)
C_e^{1N}C_h^{N1}
+O(\Gamma^6).
\end{aligned}
\label{eq:delta-detS}
\end{equation}
Substituting Eqs.~\eqref{eq:g-cross} and \eqref{eq:corner-C} yields
\begin{equation}
\delta_\theta\det S
=
\frac{
8(-1)^{N-1}\Gamma^4t^4
\bigl(\prod_jt_j^2\bigr)\cos\theta
}{E^2(E^2-4t^2)^2}
+O(\Gamma^6).
\end{equation}
The full characteristic determinant restores the two-site sweet-point
contact factors
$D_L(E)=D_R(E)=E^2(E^2-4t^2)$,
\begin{equation}
P_{\rm full}(E)=D_L(E)D_R(E)\det S(E),
\end{equation}
and therefore
\begin{equation}
\delta_\theta P_{\rm full}
=
8(-1)^{N-1}
\Gamma^4t^4
\Bigl(\prod_jt_j^2\Bigr)
E^2\cos\theta
+O(\Gamma^6).
\end{equation}
Removing the junction-independent outer Majorana factor by
$P=P_{\rm full}/E^2$ gives the common identity
\begin{equation}
\delta_\theta P(E,\theta)
=
8(-1)^{N-1}
\Gamma^4t^4
\Bigl(\prod_jt_j^2\Bigr)
\cos\theta
+O(\Gamma^6),
\label{eq:deltaP}
\end{equation}
or equivalently
\begin{equation}
B
=
8(-1)^{L_M-1}t^4\prod_{j=1}^{L_M-1}t_j^2.
\label{eq:B-lemma}
\end{equation}
Thus $\partial_E B=0$ and $\partial_{\mu_{\rm M}}B=0$ follow from the
cancellation of the lead denominators against $D_LD_R$ together with
the $E$-independence of the serial cofactors. The current still
depends on $\mu_{\rm M}$ through the spectral factor $Q$ below. The
scalar propagator $g_N(z)$ defined in Eq.~\eqref{eq:gN-body}
corresponds to the hole-block resolvent $M_h^{-1}(z)$. The
electron-block resolvent $M_e^{-1}(z)$ instead has poles at
$-\xi_q$, with the corner sign of Eq.~\eqref{eq:corner-C}. Their
end-to-end propagator moduli agree on the imaginary axis, leaving
the fourth-order kernels unchanged. Direct extraction from the full
finite Nambu matrix for these contact-reduced leads at $t=1$ confirms
$\det(E-H_{\theta=0})-\det(E-H_{\theta=\pi})
=16(-1)^{L_M-1}\Gamma^4E^2$
for $L_M=1,\ldots,6$ at representative nonzero $\mu_{\rm M}$, $E$, and
$\Gamma$, in agreement with Eq.~\eqref{eq:deltaP} after stripping
$E^2$. The identity used below is stated only through $O(\Gamma^4)$.

\paragraph{Rigorous low--high spectral windows.}
Write the single-leg channel as
$H_{\rm ch}(\theta,\Gamma)=H_{\rm ch}(\theta,0)+V_\Gamma$, where
$V_\Gamma$ contains only the two interface hoppings. For the serial
geometry $\|V_\Gamma\|\le\sqrt2\,|\Gamma|$ (operator norm). Weyl's
inequality then gives
$|E_n(\Gamma)-E_n(0)|\le\sqrt2\,|\Gamma|$ for every ordered
eigenvalue, without requiring closed-form spectra. Define
\begin{equation}
\Lambda_{\rm off}
=
\min\bigl(2t,\min_q|\xi_q|\bigr),
\qquad
\Lambda_q
=
\min\bigl(2t,\min_{p\neq q}|\xi_p|\bigr).
\end{equation}
Off resonance ($\xi_q\neq0$ for all $q$), the sufficient window
\begin{equation}
\sqrt2\,|\Gamma|
<
t_\perp
<
\Lambda_{\rm off}-\sqrt2\,|\Gamma|
\label{eq:weyl-off}
\end{equation}
guarantees that Majorana-derived low branches remain below $t_\perp$
while detuned normal and superconducting finite-energy levels remain
above it, for every $\theta$. At an isolated resonance $\xi_q=0$,
\begin{equation}
\sqrt2\,|\Gamma|
<
t_\perp
<
\Lambda_q-\sqrt2\,|\Gamma|
\label{eq:weyl-on}
\end{equation}
likewise guarantees collective clamping of the resonant low subspace
and leaves the complementary high sector unclamped. These bounds are
conservative. The main-text thresholds
$t_\perp>\sqrt2\,\Gamma_q$ (and their off-resonant analogues) are
sharper estimates from the isolated-mode effective theory. At finite
$\Gamma$ they should not be read as uncorrected exact edges. For the
arbitrary-$L_M$ identities below we use
Eqs.~\eqref{eq:weyl-off}--\eqref{eq:weyl-on} as rigorous sufficient
conditions. Broader practical windows remain as in the main text.

Within either sufficient window, a contour $|E|=t_\perp$ separates
the low eigenvalues of the single-leg channel from the high ones for
all $\theta$. Its spectral projection is analytic in $\Gamma$ near
zero. Define $L$ as the monic polynomial of the enclosed eigenvalues
and $Q$ as that of the remaining eigenvalues. Particle--hole symmetry
makes both polynomials even in $E$; changing the sign of all normal
fermions sends $\Gamma\to-\Gamma$, so their coefficients are even
in $\Gamma$. This spectral partition fixes the factors uniquely.
At $\Gamma=0$, $L_0=E^m$ with $m=2$ off resonance and $m=4$ on
resonance, while $Q_0(0)\ne0$, so the two unperturbed factors are
coprime. A process sensitive to the relative Josephson phase must
visit both interfaces, hence first occurs at order $\Gamma^4$.
Consequently, the second-order coefficients of both factors are
phase independent, and fourth-order phase matching reduces to
$B=\ell(E)Q_0(E)+E^m q(E)$.
Since the monic low factor has fixed degree $m$, its correction
$\ell(E)$ is even and has degree at most $m-2$. Regularity of $q(E)$
at zero uniquely fixes $\ell(E)$ to the Taylor polynomial of
$B/Q_0(E)$ through that degree. Thus the subtractions below follow
from the clamped spectral partition, rather than a choice of
regularization.

\paragraph{Off-resonant selective clamping.}
When every $\xi_p\neq0$,
\begin{equation}
\begin{aligned}
P_{0,\rm off}(E)
&=
E^2Q_{\rm off}(E),
\\
Q_{\rm off}(E)
&=
(E^2-4t^2)^2\prod_p(E^2-\xi_p^2).
\end{aligned}
\label{eq:P0-off}
\end{equation}
Write $P=LQ$ with monic even $L_0=E^2$. The fourth-order phase
correction of $L$ can only be a constant,
$\delta_\theta L=\Gamma^4\ell_0\cos\theta$, while
$\delta_\theta Q=\Gamma^4 q(E)\cos\theta$. Matching
$\delta_\theta P=\Gamma^4 B\cos\theta$ gives
$B=\ell_0 Q_{\rm off}+E^2 q$. The coefficients assigned to the
low-energy factor are determined by requiring the complementary factor
to remain regular at $E=0$. With
$A_{\rm off}=B/Q_{\rm off}$ and $Q_{\rm off}(0)\neq0$ one finds
$\ell_0=A_{\rm off}(0)$ and
\begin{equation}
\delta_\theta\ln Q(E)
=
\Gamma^4\cos\theta\,
\frac{A_{\rm off}(E)-A_{\rm off}(0)}{E^2}
+O(\Gamma^6).
\label{eq:lnQ-off}
\end{equation}
This Laurent decomposition assigns the singular piece at $E=0$ to the low-energy factor, leaving a regular high-sector weight that matches the BdG partition inside the window of Eq.~\eqref{eq:weyl-off}. On the Matsubara axis,
\begin{equation}
A_{\rm off}(i\omega)
=
-\frac12
\left(\frac{4t^2}{\omega^2+4t^2}\right)^2
|g_N(i\omega)|^2.
\end{equation}
In the smooth single-clamped window the $O(\Gamma^2)$ clamped branch
carries no phase dependence. The unclamped high sector of the ladder
obeys
\begin{equation}
E_{\rm GS}^{\rm high}
=
-\frac1\pi\int_0^\infty\ln Q(i\omega)\,d\omega
+\mathrm{const},
\end{equation}
where $1/\pi$ (rather than $1/(2\pi)$ for a single-leg sum
$-\frac12\sum_a E_a$) accounts for the two helicity sectors. With
$E_{\rm GS}=-J_1\cos\theta+\cdots$ one recovers Eq.~\eqref{eq:J1-GF}.
Because $0<(4t^2/(\omega^2+4t^2))^2<1$ and
$|g_N(i\omega)|^2<|g_N(0)|^2$ for $\omega>0$, the integrand
is negative and $J_{1,\rm off}^{(4)}<0$ for any finite serial $L_M$
whenever this kernel applies. Off resonance the detuned normal levels
remain inside $Q_{\rm off}$ and still participate in phase selection.
For $L_M=1$ and $\varepsilon=\mu_{\rm M}$ the integral evaluates to
\begin{equation}
J_{1,\rm off}^{(4)}
=
-\Gamma^4
\left[
\frac{1}{4|\varepsilon|^3}
+
\frac{8t+3|\varepsilon|}
{16t|\varepsilon|(2t+|\varepsilon|)^2}
\right],
\label{eq:app-J1-off-full}
\end{equation}
whose first term is the end-mode Laurent piece and whose second is a
strictly negative finite-energy correction. Near an isolated
detuned mode $q$, the leading low-energy contribution is obtained by
replacing $(\varepsilon,\Gamma)$ with $(\xi_q,\Gamma_q)$. The
complete finite-segment coefficient, including interference among
normal modes, is given by Eq.~\eqref{eq:J1-GF}. The detuning--threshold relation
Eq.~\eqref{eq:threshold-relations}
and the explicit scales of
Eq.~\eqref{eq:app-three-scales}
remain the working asymptotic estimates. The rigorous sufficient
window is Eq.~\eqref{eq:weyl-off}.

\paragraph{Resonant collective clamping.}
At $\phi=\pi$,
$[H_\perp,I_{\rm leg}\otimes H_{\rm ch}^{(L_M)}]=0$, so
Eq.~\eqref{eq:maxformula} is exact for any $L_M$. When $\xi_q=0$ the
resonant normal level enters the low-energy subspace and
\begin{equation}
\begin{aligned}
P_{0,\rm on}(E)
&=
E^4Q_q(E),
\\
Q_q(E)
&=
(E^2-4t^2)^2\prod_{p\neq q}(E^2-\xi_p^2).
\end{aligned}
\label{eq:P0}
\end{equation}
Inside the sufficient window~\eqref{eq:weyl-on} the resonant low
pair is clamped ($J_{1,\rm low}=0$) while higher states remain
unclamped. The main-text interval
$\sqrt2\,\Gamma_q<t_\perp<\Lambda_q$ is the corresponding sharper
asymptotic estimate. The monic even factor $L_0=E^4$ admits constant
and $E^2$ fourth-order phase corrections. Matching coefficients in
$P=LQ$ with $A_q=B/Q_q$ and $Q_q(0)\neq0$ gives $\ell_0=A_q(0)$
and the quadratic matching coefficient $\frac12 A_q''(0)$. Then
\begin{equation}
\delta_\theta\ln Q(E)
=
\Gamma^4\cos\theta\,
\frac{
A_q(E)-A_q(0)-\frac{E^2}{2}A_q''(0)
}{E^4}+O(\Gamma^6).
\label{eq:lnQ}
\end{equation}
Analogously, the two Laurent subtractions at resonance absorb the singular pieces into the quartic low factor, leaving the regular high-sector remainder. On the
Matsubara axis, the regular high-energy contribution is described by
$R_q(\omega)$ defined in Eq.~\eqref{eq:Rq-main}. Its product
representation establishes strict convexity in $\omega^2$, yielding,
for $\omega>0$,
\begin{equation}
R_q(\omega)-R_q(0)-\frac{\omega^2}{2}R_q''(0)>0,
\label{eq:Rq-pos}
\end{equation}
hence $J_{1,\rm on}^{(4)}>0$. Equivalently,
$R_q(\omega)=\frac{\omega^2}{2}(4t^2/(\omega^2+4t^2))^2
|g_N(i\omega)|^2$ at $\xi_q=0$. The resonant level no longer
belongs to $Q_q$. It has been moved into the clamped low-energy
factor. The same high-sector integral with prefactor $1/\pi$ produces
Eq.~\eqref{eq:J1-res}.

\emph{Consistency check ($L_M=1$).} With $\psi_1(1)=1$,
$R_1(\omega)=\frac12(4t^2/(\omega^2+4t^2))^2$,
$\frac12 R_1''(0)=-1/(4t^2)$, and
\begin{equation}
J_1^{(4)}
=
\frac{\Gamma^4}{\pi}
\int_0^\infty d\omega\,
\frac{
\frac12\left(\frac{4t^2}{\omega^2+4t^2}\right)^2
-\frac12+\frac{\omega^2}{4t^2}
}{\omega^4}
=
\frac{5}{64}\frac{\Gamma^4}{t^3},
\label{eq:5-64-check}
\end{equation}
recovering Eq.~\eqref{eq:J1-eq37}.

\FloatBarrier

\section{Representative Resonant CPRs}
\label{app:rep-cpr}

\begin{figure}[htbp]
\centering
\includegraphics[width=\columnwidth]{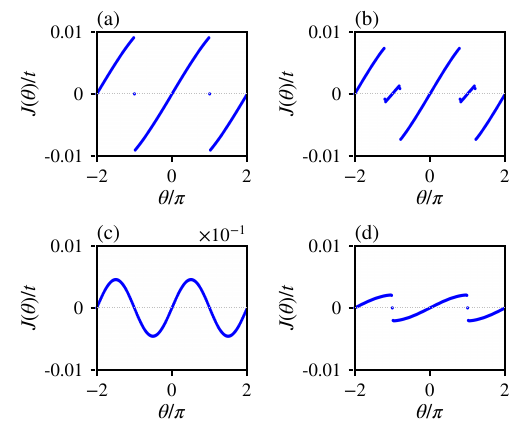}
\caption{Current-phase relations for $L_{\rm M}=5$ and $\mu_{\rm S}=\mu_{\rm M}=t$, with $\Delta/t=0.6$, $\phi=\pi/3$, $\Gamma/t=0.1$, and $T/t=0.0001$, as in Sec.~\ref{results}. Panels (a)--(d) correspond to $t_\perp/t=0$, $0.02$, $0.1$, and $1.2$, respectively.}
\label{fig:J50}
\end{figure}

Figure~\ref{fig:J50} shows representative resonant CPRs for increasing
$t_\perp$. The decoupled ($t_\perp=0$) and weakly coupled
($t_\perp=0.02t$) cases retain topological/mixed branch switching.
The intermediate resonant case ($t_\perp=0.1t$) exhibits the smooth
$0$-type response discussed in Sec.~\ref{results}. At stronger
inter-leg coupling ($t_\perp=1.2t$) the CPR is changed again
without restoring the off-resonant $\pi$ component. These panels
complement the level-alignment diagnostics of
Fig.~\ref{fig:resonance-diag}.

\begingroup
\hypersetup{urlcolor=black}
\bibliographystyle{jpsj_doi}
\bibliography{Reference_JPSJ}
\endgroup

\end{document}